\documentclass[12pt]{article}
\usepackage{latexsym}
\usepackage{amssymb}
\usepackage{amsfonts}
\usepackage{pslatex}
\usepackage{graphicx}

\catcode`\@=11
\def\numberbysection{\@addtoreset{equation}{section}
        \def\theequation{\thesection.\arabic{equation}}}

\def\be{\begin{equation}}
\def\ee{\end{equation}}
\def\ba{\begin{eqnarray}}
\def\ea{\end{eqnarray}}

\def\sc{\scriptsize}

\def\ov{\overline}

\def\Z{\mathbb{Z}}

\def\nl{\nonumber \\}

\def\ra{\rangle}

\def\la{\langle}
\def\de{\partial}
\def\wt{\widetilde}

\def\Tr{{\rm Tr}}
\def\dag{\dagger}

\def\a{\alpha}
\def\b{\beta}
\def\g{\gamma}
\def\G{\Gamma}
\def\D{\Delta}
\def\d{\delta}

\def\eps{\varepsilon}

\def\th{\theta}

\def\l{\lambda}
\def\L{\Lambda}

\def\P{\Pi}
\def\r{\rho}
\def\s{\sigma}

\def\w{\omega}

\def\B{{\bf B}}
\def\ap{\approx}

\numberbysection

\begin{document}
\begin{titlepage}
\begin{center}

\hfill  \quad DFF 438/11/07 \\

\vspace{1cm} 

{\Large \bf Semiclassical Droplet States } \\ 
\vspace{.4cm}
{\Large \bf in Matrix Quantum Hall Effect }\\ 

\vspace{1cm}

Andrea CAPPELLI ${}^a$\ ,\ \ Ivan D. RODRIGUEZ ${}^{a,b}$ \\
\bigskip
${}^a$ {\em Istituto Nazionale Fisica Nucleare}\\
${}^b$ {\em Dipartimento di Fisica, Universit\'a di Firenze}\\
{\em  Via G. Sansone 1, 50019 Sesto Fiorentino (FI), Italy} \\
\end{center}
\vspace{.5cm}
\begin{abstract}
We derive semiclassical ground state solutions that correspond to the
quantum Hall states earlier found in the Maxwell-Chern-Simons
matrix theory. They realize the Jain composite-fermion construction
and their density is piecewise constant as that of phenomenological
wave functions. These results support the matrix theory as a
possible effective theory of the fractional Hall effect.
A crucial role is played by the constraint limiting the degeneracy
of matrix states: we find its explicit gauge invariant form
and clarify its physical interpretation.
\end{abstract}

\vfill
\end{titlepage}
\pagenumbering{arabic}


\section{Introduction}

In this paper we continue the study
of noncommutative and matrix gauge theories 
\cite{NCFT}\cite{ncgeom}\cite{mtheory}
that can be used as effective non-relativistic theories\footnote{
We refer to \cite{cr} for an introduction
to noncommutative theories of the quantum Hall effect.}
\cite{fradkin}\cite{cftheories} of the
fractional Hall effect \cite{prange}.
The original Susskind proposal was based on the noncommutative 
Chern-Simons theory \cite{susskind}: it
developed in the Polychronakos Chern-Simons matrix model
\cite{poly1}, that was analysed by several authors 
\cite{poly3} \cite{heller} \cite{karabali}
\cite{hansson} \cite{hansson2} \cite{leigh} \cite{cr}\cite{lambert}.
This model is a $U(N)$ gauge theory in one dimension (i.e. time) 
with Chern-Simons kinetic term; it involves two Hermitean 
matrix coordinates,
$X_1(t),X_2(t)$, that are noncommuting, $[X_1,X_2]=i\ \th \ I$,
where $\th$ is a constant ``background charge'' and $I$ is the identity matrix.
In the semiclassical limit \cite{susskind}, the matrix model describes
two-dimensional incompressible fluids in strong magnetic field $\B$ with
Laughlin's values of the filling fraction, $\nu=1/(k+1)$, where 
$k=\B\th$ is integer quantized \cite{nair}.

At the full quantum level, some problems were found in 
matching the matrix model to the physics of the fractional Hall effect:
one point was that the theory reduced to the eigenvalues, 
$\l_a$, $a=1,\dots,N$,
of $X=X_1+i\ X_2$, does not really describe electrons in the lowest 
Landau level with coordinates $\l_a$; owing to matrix noncommutativity,
the Laughlin wave function is deformed at 
short distances  \cite{karabali}\cite{cr}.
Another issue was that this model cannot easily describe the general 
Hall states with Jain's filling fractions: 
$\nu=m/(mk\pm 1)$, $m=2,3,\dots$ \cite{jain}.

In our earlier paper \cite{c-rod},  
we showed that these problems can be overcome by 
upgrading the Chern-Simons model to the Maxwell-Chern-Simons matrix theory.
This includes an additional kinetic term quadratic in time derivatives and
the potential, $V = - g \Tr \left[ X_1,X_2 \right]^2$, with the 
coupling $g \ge 0$ that controls matrix noncommutativity.
All terms in the action are determined by the gauge principle
because they can be obtained by dimensional reduction of
Maxwell-Chern-Simons gauge theory, as 
discussed in the literature of D0-branes in string theory \cite{mtheory}.

The Maxwell-Chern-Simons matrix theory reduces to the earlier Chern-Simons
model for large  values of $\B$ (with finite $g$).
However, in the different $g=\infty$ limit (with finite $\B$), corresponding
to $[X_1,X_2]=0$, it does provide a sensible physical description
of the fractional Hall effect: after reduction to eigenvalues, one finds
electrons in Landau levels interacting  
with a two-dimensional potential of Calogero type, $1/|\l_a-\l_b|^2$ 
(see section 2.2). 
For general $\B<\infty$, the additional coupling $g$ in the theory allows
one to interpolate between matrix ($g=0$) and electron 
($g=\infty$) dynamics.

Furthermore, the theory is exactly solvable at $g=0$ \cite{c-rod}: 
it was found to
describe a matrix extension of the Landau levels, where the gauge-invariant
many-body states are given by matrix generalizations of Slater determinants. 
Although the degeneracy of matrix states grows exponentially with energy, 
it was possible to control it by introducing suitable projections 
in the theory.
We showed that the constraint $\left(A_{ab} \right)^m \approx 0 $, 
$\forall a,b$, projects the theory to the ``lowest $m$ matrix Landau levels'',
with $m=1,2,\dots$; the $m$-th reduced theory naturally possesses 
non-degenerate homogeneous ground states with filling fractions of the
Jain series, $\nu=m/(mk+1)$ \cite{c-rod} (see section 2). 
Indeed, the solutions of the gauge invariance conditions 
and of the constraint $A^m \ap 0$ give raise to a rather surprising structure 
of ground states that corresponds to 
the Jain composite-fermion construction of ansatz wave functions \cite{jain}.

These $g=0$ states  exactly match the phenomenological Jain
wave functions under matrix diagonalization, that is formally
achieved at  $g=\infty$.
Therefore, we conjectured that these matrix ground states 
have smooth deformations for $g>0$ into the physical $g=\infty$ states, 
namely that no phase transitions are found for  $0<g<\infty$ when
the system is at the specific densities \cite{c-rod}.

A number of problems remained to be further investigated:
\begin{itemize}
\item
Understand the matrix Jain states, e.g. compute their densities 
and observables quantities.
\item
Understand the projections  $\left(A_{ab} \right)^m \ap 0 $ 
in more physical terms.
\item
Study the phase diagram of the theory for $0<g<\infty$ at the relevant 
densities.
\end{itemize}

In the present paper, after reminding earlier results \cite{c-rod}
(section 2), we find the gauge invariant form of the projection $A^m\ap 0$
 and its semiclassical physical meaning in terms of single-particle
 occupancy (section 3). Next, we study the matrix Jain states
in the semiclassical approximation, by analytically solving
the classical equations of motion, further constrained by the Gauss law 
and the semiclassical version of the $A^m \ap 0$ condition (section 4).
The ground states are found to be two-step
droplets of incompressible fluid with piecewise constant density;
this is the same density shape of the phenomenological Jain states 
before projection to the lowest Landau level \cite{jain} (where the
density of incompressible fluids becomes strictly constant).

The fact that the matrix Jain states at $g=0$ already
have the expected droplet density of physical $g=\infty$ states,
supports our earlier claim that these ground states could
remain stable while varying $0<g<\infty$ \cite{c-rod}.
Other ground states corresponding to generalized Jain constructions
with different filling fraction, although possible in the $g=0$ theory, 
are found not to possess piecewise constant density.
We argue that the modulated density shape is a signal of 
ground-state instability at finite $g$ values, since the
corresponding phenomenological Jain states ($g=\infty$) 
are known to be unstable \cite{jain}.
We complete our study of semiclassical solutions by describing the 
quasi-holes excitations above the matrix Jain states.
Finally, in the conclusion we briefly discuss the ways to study the
Maxwell-Chern-Simons matrix theory for $g>0$.


\section{Jain states in Maxwell-Chern-Simons matrix gauge theory}

\subsection{Lagrangian and Hamiltonian}

In this section we recall the matrix theory of 
quantum Hall states proposed in \cite{c-rod} 
and the derivation of ground states 
that are matrix analogs of the Jain composite-fermion states.
The theory involves three time-dependent $N\times N$ Hermitean matrices,
$X_i(t)$, $i=1,2$ and $A_0(t)$, and an auxiliary complex vector $\psi(t)$.
The Lagrangian contains a Maxwell-Chern-Simons kinetic term,  
a uniform ``charge background'' $\th$ and the $\psi$ 
``boundary term'' of ref.\cite{poly1}:
\ba
S&=&\int \ dt\ \Tr\left[
\frac{m}{2}\left(D_t\ X_i \right)^2 \ +\ 
\frac{\B}{2} \eps_{ij} \ X_i\ D_t \ X_j \ +\ 
\frac{g}{2} \left[ X_1,X_2\right]^2 \ +\ \B\th\ A_0
\right]
\nl
& &-i\ \int \psi^\dag\ D_t \psi \ .
\label{mcs-action}
\ea
The covariant derivatives are:
$D_t X_i=\dot{X}_i -i \left[A_0,X_i\right]$ and
$D_t \psi = \dot{\psi}-i A_0\psi$.
Under the $U(N)$ gauge transformations: $X_i\ \to\ U X_i U^\dag$,
$A_0\ \to\ U \left(A_0-id/dt \right) U^\dag$, 
and $\psi\ \to\ U\psi$, the action changes by
a total derivative, such that invariance under large gauge
transformations requires the quantization,  $\B\th=k\in \Z$ \cite{nair}.
Hereafter we set $m=1$ and measure dimensionful constants accordingly.

The variation of $S$ w.r.t. the non-dynamical field $A_0$ gives the 
Gauss-law constraint; its expression in term of coordinates $X_i$ 
and conjugate momenta $\P_i$, $i=1,2$, reads:
\be
G\ \ap \ 0\ , \qquad\quad
G\ =\ i\ \left[X_1,\P_1\right]\ +\ i\ \left[X_2,\P_2\right]\ 
-\ \B\th\ {\rm I}\ +\ \psi\otimes \psi^\dag \ .
\label{mcs-gauss}
\ee
The trace of $G$ fixes the norm of the auxiliary vector $\psi$,
\be
\Tr\ G\ =\ 0\quad \longrightarrow\ \Vert\psi\Vert^2 =\B\th N=kN\ .
\label{psi-norm}
\ee
We note that $\psi$ has 
trivial dynamics, $\psi(t)=\psi(0)={\rm const.}$, and it is necessary
to represent the Gauss law on
finite-dimensional matrices that have traceless commutators \cite{poly1}.
In a gauge in which all $\psi$ components vanish but the last one,
the term $\left( k {\rm I}\ -\psi\otimes \psi^\dag \right)$ 
in (\ref{mcs-gauss}) becomes the ``traceless identity'',
$k \ {\rm I}_N$, $\ {\rm I}_N=diag(1,\cdots,1,1-N)$.

In gauge-invariant quantization, all $2N^2$ matrix degrees of freedom 
$X^i_{ab}$ are quantized and the Gauss law is imposed on states:
$G$  generates 
$U(N)$ gauge transformations of $X_i$ and $\psi$, and $G = 0$ 
implies that physical states are $U(N)$ singlets subjected
to the additional condition (\ref{psi-norm}) fixing the total number of
$\psi_a$ components equal to $kN$.
The Hamiltonian can be written:
\be
H \ =\ \B\ \Tr \left( A^\dag \ A \right)\ +\ \frac{\B}{2} N^2\ -
\frac{g}{2} \Tr \left[X_1,X_2 \right]^2\ , 
\label{a-ham}
\ee
after introducing the variable,
\be
A=\frac{1}{2\ell}\left( X_1 +i\ X_2 \right)\ +\
\frac{i\ell}{2}\left( \P_1 +i\ \P_2 \right)\ ,
\qquad 
\label{a-def}
\ee
and its adjoint $A^\dag$, involving the magnetic length,
$\ell=\sqrt{2/\B}$.
These quantities obey the commutation relations of $N^2$ 
independent harmonic oscillators:
\be
 \left[\left[A_{ab},A^\dag_{cd} \right]\right] \ = \ \d_{ad}\ \d_{bc}\ \ ,
\qquad\qquad
\left[\left[A_{ab},A_{cd}\right]\right] \ =\  0\ .
\label{a-comm}
\ee
Therefore, for $g=0$ the Hamiltonian describes Landau levels
populated by $N^2$ two-dimensional ``particles'' with phase-space coordinates,
$\{\P^i_{ab},X^i_{ab}\}$, $a,b=1,\dots,N$, $i=1,2$.
Degenerate angular momentum excitations are described 
by an independent set of oscillators:
\be
B=\frac{1}{2\ell}\left( X_1 -i\ X_2 \right)\ +\
\frac{i\ell}{2}\left( \P_1 -i\ \P_2 \right)\ ,
\qquad 
\label{b-def}
\ee
\be
 \left[\left[B_{ab},B^\dag_{cd} \right]\right] \ = \ \d_{ad}\ \d_{bc}\ \ ,
\qquad\qquad
\left[\left[B_{ab},B_{cd}\right]\right] \ =\  0\ ,
\label{b-comm}
\ee
that commute with $A,A^\dag$.
The total angular momentum of the $N^2$ particles is
\be
J\ =\ \Tr\left( X_1\ \P_2\ - \ X_2\ \P_1 \right)\ = \
 \Tr\left( B^\dag B\ - \ A^\dag A\right)\ .
\label{j-def}
\ee

The $N^2$- particle states are further constrained by the Gauss law 
(\ref{mcs-gauss}): as we shall see later, this enforces a kind of generalized
exclusion statistics.
It is convenient to introduce the complex matrices,
\ba
X &=& X_1+i\ X_2\ , \qquad\qquad \ov{X}=X_1 -i\ X_2\ ,
\nl
\P &=& \frac{1}{2}\left(\P_1 -i\ \P_2\right)\ ,
\qquad \ov{\P}=\frac{1}{2}\left(\P_1 +i\ \P_2\right)\ ,
\label{b-mat}
\ea
and use the bar for denoting the Hermitean conjugate of
classical matrices, keeping the dagger for the quantum adjoint.
Hereafter we set the magnetic length to one, i.e. $\B=2$.

For states with constant density\footnote{
See section 4 for the definition of density in matrix theories.},
the angular momentum measures the extension of the ``droplet of fluid'',
such that we can relate it to the semiclassical
filling fraction $\nu_{\rm cl}$ by the formula,
\be
\nu_{\rm cl}\ =\ \lim_{N\to\infty}\ \frac{N(N-1)}{2 \la J\ra}\ .
\label{nu-def}
\ee

In a physical system of finite size, one can control
the density of the droplet by adding
a confining potential $V_C$ to the Hamiltonian:
\be
H \ \to \ H\ +\ V_C \ =\ H\ +\ \w\ \Tr\left(B^\dag\ B  \right) \ 
+\ \w_n\ \Tr\left(B^{\dag n}\ B^n  \right)\ .
\label{conf-pot}
\ee
The strength $\w$ is of order $O(\B/N)$ such that the structure of
Landau levels is not destroyed by putting  $n_B=O(N)$ particles per level.
The higher order terms $O(\w_n)$ also commute with the $g=0$
Hamiltonian: one can show that their eigenvalues on constant-density states
are very large $O(N^{n-1})$ for fillings
$n_B>n$ and thus they can be used to
simulate finite-box boundary conditions \cite{c-rod}.

\begin{figure}
\begin{center}
\includegraphics[width=10cm]{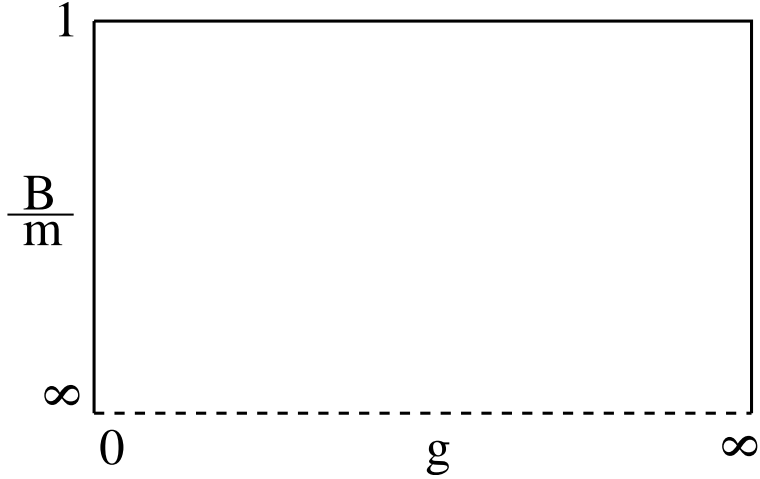}
\end{center}
\caption{Phase diagram of the Maxwell-Chern-Simons matrix theory.
The vertical axes $g=0$ and $g=\infty$ have been discussed 
in ref. \cite{c-rod}.
The Chern-Simons matrix model \cite{poly1} is found at $B\to\infty$ in 
the left down corner.}
\label{phase}
\end{figure}

In Figure (\ref{phase}) we illustrate the phase diagram of the
Maxwell-Chern-Simons matrix theory as a function of the
parameters $\B/m$ and $g$.
The analysis of \cite{c-rod} found the properties of the theory 
on the axes $g=0$ and $g=\infty$. For $g=0$, the theory is exactly
solvable and possess a set of ground states that are in 
one-to-one relation with the Laughlin and Jain ground states
with filling fractions $\nu=m/(mk+1)$. 
These states are selected by the value of
$k$ and by adjusting the density via the parameters $\w,\w_n$ of 
the confining potential (\ref{conf-pot}); furthermore, 
they are unique non-degenerate states obeying the conditions
$\left(A_{ab}\right)^m\Psi=0$
that projects out degeneracies specific of matrix states, as it 
will be explained momentarily.

The limit $\B\to\infty$ at $g=0$ (and $g$ finite) in the  Hamiltonian 
(\ref{a-ham}) corresponds to vanishing kinetic term, i.e. to
the constraints $A=A^\dag = 0$ of lowest Landau level \cite{dunne}.
In this limit, the Gauss law (\ref{mcs-gauss}) reduces to:
\be
G \ = \  - i\B\ \left[X_1,X_2\right] - \B\th +\psi\otimes\psi^\dag\ ,
\label{cs-gauss}
\ee
and it uniquely fixes the noncommutativity of matrices. 
The potential term in $H$ (\ref{a-ham}) becomes a constant 
for all physical states
and the theory reduces to the Chern-Simons matrix model \cite{poly1}
with trivial dynamics, $H=V_C$ \cite{c-rod}.

For $g=\infty$ ($\B$ finite),  the theory describes $N$ ordinary 
electrons with coordinates
given by the complex eigenvalues  $\l_a$, $a=1,\dots,N$, of $X$, 
interacting with the two-dimensional Calogero potential.

\subsection{$g\to\infty$ limit and electron theory}

For large $g$ values, the potential term in $H$ (\ref{a-ham}), 
$g\ \Tr [X,\ov{X}]^2$, restricts
the configuration space of complex matrices to those commuting with
their conjugate, the so-called ``normal'' matrices \cite{wiegmann}.
Therefore, $X$ can be made diagonal by unitary (gauge) transformation
and the theory completely reduces to the eigenvalues, that are interpreted
as electron coordinates.
Any complex matrix can be written \cite{mehta} as, $X=\ov{U}(\L+R)U$, 
in terms of a unitary matrix $U$ 
(the gauge degrees of freedom), a diagonal matrix $\L$ (the eigenvalues) 
and a upper triangular complex matrix $R$ (additional d.o.f.). 
Therefore, the gauge-invariant degrees of freedom different from the
eigenvalues contained in the $R$ matrix are suppressed for
$g\to\infty$.

We can thus take the diagonal gauge for $X$
and decompose the momenta $\Pi,\ov{\Pi}$
in diagonal and off-diagonal matrices, respectively called $p$  and $\G$:
\be
X=\L \ , \qquad \P=p+\G\ ,\quad \ov{\P}=\ov{p}+\ov{\G}\ .
\label{l-expr}
\ee
In this gauge, the Gauss law (\ref{mcs-gauss}) can be solved at the classical 
level and it determines the off-diagonal momenta,
\be
\G_{ab}\ =\ \frac{i k}{2}\ 
\frac{\ov{\l}_a-\ov{\l}_b}{\vert \l_a-\l_b\vert^2}\ ,
\qquad\quad a\neq b\ .
\label{gamma-val}
\ee
Upon inserting them into the Hamiltonian (\ref{a-ham}), the
diagonal and off-diagonal components decouple and one obtains,
\be
H\vert_{g=\infty}\ =\ 2\ \sum_{a=1}^N \ \left(\frac{\ov{\l}_a}{2} -ip_a\right)
\left(\frac{\l_a}{2} +i\ov{p}_a\right) \ +\ 
\frac{k^2}{2}\ \sum_{a \neq b =1}^N\ 
\frac{1}{\vert \l_a-\l_b\vert^2}\ .
\label{inf-ham}
\ee
The quantization can be done on the remaining independent variables, 
which are the complex eigenvalues $\l_a$ and their conjugate momenta
$p_a$ \cite{poly97}.
Therefore, the theory reduced to the eigenvalues corresponds
to the ordinary Landau problem of $N$ electrons plus an induced
two-dimensional Calogero interaction.
The measure of integration on matrices
(\ref{int-meas}) also reduces to that of ordinary electrons
after incorporating one Vandermonde factor in the wave
functions \cite{wiegmann}. This causes a renormalization 
of the filling fraction from the semiclassical
expression (\ref{nu-def}): $1/\nu=1+ 1/\nu_{\rm cl}$

We conclude that the Maxwell-Chern-Simons matrix theory in the
$g=\infty$ limit makes contact with the physical problem 
of the fractional quantum Hall effect: the only difference
is that the Coulomb repulsion $e^2/r$  
is replaced by the Calogero interaction $k^2/r^2$.
Numerical results \cite{laugh}\cite{haldane} \cite{jain}\cite{numeric}
indicate that quantum Hall incompressible
fluid states are rather robust and do not depend on the detailed form of the
repulsive potential at short distance, for large $\B$, at least for the
qualitative features.
The $g=\infty$ theory is not, by itself, less
difficult than the ab-initio quantum Hall problem: the gap is
non-perturbative and there are no small parameters.
The advantage of embedding the problem into the matrix theory is 
that of making contact with the solvable $g=0$ limit.


\subsection{Matrix Jain states at $g=0$}

The wave functions of the Maxwell-Chern-Simons theory 
take the form,
\be
\Psi\ = \ e^{-\Tr\left(\ov{X} X \right)/2 -\ov{\psi} \psi/2}\ 
\Phi(X,\ov{X},\psi) \ ,
\label{wf-def}
\ee
and their integration measure reads:
\be
\left\la \Psi_1 \vert \Psi_2 \right\ra \ = \
\int {\cal D}X {\cal D}\ov{X}\ {\cal D}\psi {\cal D}\ov{\psi}\  
e^{-\Tr \ov{X} X -\ov{\psi} \psi}\ 
\Phi^*_1(X,\ov{X},\psi)\ \Phi_2(X,\ov{X},\psi)\ .
\label{int-meas}
\ee
At $g=0$, the energy and momentum eigenstates are obtained by applying 
powers of the $A^\dag_{ab}$ and $B^\dag_{ab}$ operators 
(\ref{a-def},\ref{b-def}) to the empty ground state
$\Psi_o=\exp\left(-\Tr \ov{X}X/2 - \ov{\psi}\psi/2\right)$
(as in ordinary Landau levels), leading to:
\be
\Psi\ = \ e^{-\Tr\ \ov{X} X/2 -\ov{\psi} \psi/2}\ 
\Phi(\ov{B},\ov{A},\psi) \ ,\qquad
E=\B\ N_A\ , \quad {\cal J}=N_B-N_A\ .
\label{wf-def2}
\ee
The wavefunction $\Phi$ is a homogeneous polynomial of 
$\ov{B}=X-\de/\de\ov{X}$ and  $\ov{A}=\ov{X}-\de/\de X$, that
can be treated as $c$-number matrices because
they commute among themselves.
The energy  $E=\B N_A$ and momentum ${\cal J}=N_B-N_A$ of the state
are expressed in terms of the eigenvalues of
 $\Tr\ A^\dag A$ and $\Tr\ B^\dag B$, 
$N_A$ and $N_B$ respectively, that count the total number of matrices
$\ov{A}$ and $\ov{B}$ in $\Phi$.

In the lowest Landau level, the wave function obeys 
$A_{ab}\Psi=0$, $\forall\ a,b$, thus  the polynomial part $\Phi$
 does not contain any  $\ov{A}_{ab}$: it
is a analytic function of the $\ov{B}_{ab}$ variables,
equal to the $X_{ab}$. 
The physical states $\Phi(X,\psi)$ obeying the Gauss law are $U(N)$ singlets 
that contain any number of $X_{ab}$ and $Nk$ copies of the  $\psi$ vector
(owing to (\ref{psi-norm})).
The solutions for $k=1$ are given by \cite{heller}:
\be
\Phi_{\{n_1,\dots,n_N\}}\left(X,\psi\right)\ = \ 
\eps^{a_1\dots a_N}\ \left(X^{n_1}\psi\right)_{a_1}\cdots
\left(X^{n_N}\psi\right)_{a_N} \ ,\qquad 0\le n_1<n_2<\cdots <n_N\ ,
\label{HVR-states}
\ee
where $\eps$ is the completely antisymmetric tensor and $\{n_i\}$
any ordered set of integers.
Solutions for $k> 1$ are obtained by multiplying
$k$ terms (\ref{HVR-states}), leading to the expressions,
$\Phi_{\{n^1_1,\dots,n^1_N\}\cdots\{n^k_1,\dots,n^k_N\}}$.
The ground state in the confining potential $\Tr( X X^\dag)$
is given by the closest packing $\{0,1,\dots,N-1\}$ that has the lowest
angular momentum, i.e. lowest degree in $X$:
\be
\Phi_{k,\ gs}=\left[\eps^{a_1\dots a_N}\ \psi_{a_1} 
\left(X\psi\right)_{a_2}\cdots
\left(X^{N-1}\psi\right)_{a_N} \right]^k \ .
\label{Laugh-mat}
\ee

If we diagonalize the complex matrix $X$ by similarity transformation:
$X = V^{-1}\ \L\ V $, $ \L =  diag(\l_1,\dots,\l_N)$,
$\psi = V^{-1}\ \phi$,
the dependence on $V$ and $\phi$ factorizes 
and the powers of eigenvalues 
make up the Vandermonde determinant $\D(\l)=\prod_{a<b}(\l_a-\l_b)$,
as follows: 
\be
\Phi_{k,\ gs}\left(\L,V,\psi\right) = \left(\det V\right)^{-k} 
\ \prod_{1\le a< b\le N}\left(\l_a-\l_b\right)^k\  
\left(\prod_c \ \phi_c\right)^k\ .
\label{Laugh-wf}
\ee
The central piece is indeed the Laughlin wave function \cite{laugh}, 
upon interpreting the eigenvalues as
electron coordinates \cite{heller} \cite{karabali}. 
The value of the filling fraction (\ref{nu-def}) is:
\be
\nu\ =\ \frac{1}{k+1}\ ,
\label{nu}
\ee
by keeping into account the
extra Vandermonde coming from the integration measure.

Eq.(\ref{Laugh-wf}) 
is the most interesting result obtained in the noncommutative approach and
the Chern-Simons matrix model \cite{poly1}\cite{heller}: that of deriving
the Laughlin wave function from gauge invariance in a matrix theory
with background charge $\th$.
Furthermore, Susskind's semiclassical analysis \cite{susskind} 
showed that, in the limit $\th\to 0$,
this matrix state  describes an incompressible fluid in high magnetic fields
with density, $\rho_o\ = \ 1/(2\pi \th)$ ,
and classical filling fraction $\nu_{\rm cl}=2\pi\rho_o/\B=1/\B \th =1/k$
in agreement with the earlier identification.

The analog of the Laughlin quasi-holes are realized
by shifting the occupation numbers of matrices, e.g. by the state 
in (\ref{HVR-states}) with 
$\{n_1,n_2,\cdots,n_M\}=\{1,2,\cdots,N\}$.
This has $\D {\cal J} =O(N)$ and thus a finite gap
$\D E=O(\B)$. 
On the other hand, the quasi-particle excitations cannot be 
realized in the Chern-Simons matrix model \cite{poly1}.
In general, states with higher density do not exist in this theory, 
because they would
need to populate higher Landau levels that are absent.

Further difficulties in matching the Chern-Simons matrix model to
the Laughlin physics at the quantized level were discussed in Refs.
\cite{karabali}\cite{cr}\cite{hansson2}: 
since the matrices $X,\ov{X}$ are noncommuting
(cf. (\ref{cs-gauss})), the theory is not well suited for 
the description of electron degrees of freedom.
The reduction \cite{cr} to eigenvalues $(\l_a,\ov{\l}_a)$, interpreted
as electron coordinates, shows that
the measure of integration (\ref{int-meas}) does not become
that of the Landau levels and the Laughlin state
(\ref{Laugh-wf}) gets deformed at short distance.
These findings should be contrasted with the results in section 2.2 for
Maxwell-Chern-Simons matrix theory, that possesses
a well-definite physical limit at $g=\infty$ (although difficult to solve).

The states (\ref{HVR-states})
can be represented graphically as ``bushes'', as shown in Fig.(2a).
The matrices $\ov{B}_{ab}$ (i.e. $X_{ab}$) are depicted as oriented 
segments with indices at 
their ends and index summation amounts to joining segments
into lines, as customary in gauge theories.
The lines are the ``stems'' of the bush ending with one $\psi_a$, represented
by an open dot, and the epsilon tensor is the N-vertex
located at the root of the bush. 
Bushes have N stems  of different lengths:
$n_1 < n_2 <\cdots <n_N$. 
The position $i_\ell$ of one
$\ov{B}$ on the $\ell$-th stem, $1\le i_\ell\le n_\ell$, 
is called the ``height'' on the stem. 

\begin{figure}
\begin{center}
\includegraphics[width=10cm]{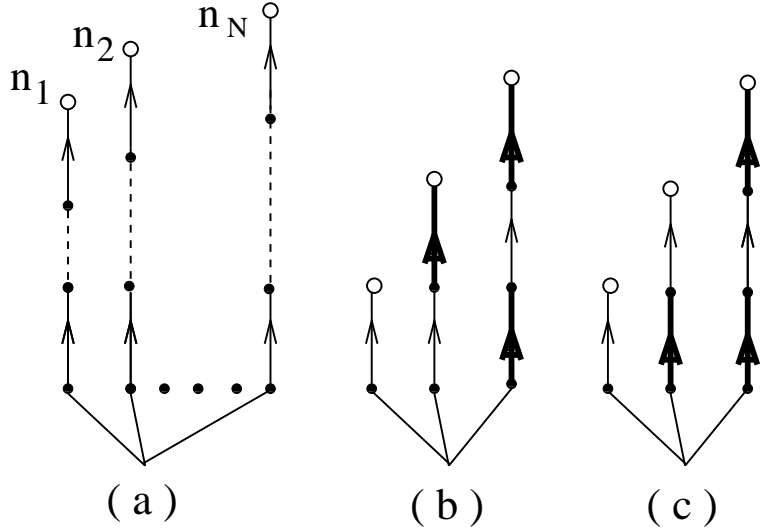}
\end{center}
\caption{Graphical representation of gauge invariant states: (a) 
general states in the lowest Landau level (cf. Eq.(\ref{HVR-states})); 
(b) and (c), $N=3$ examples in the second and third levels involving both
matrices, $\ov{B}$ (thin line) and $\ov{A}$ (in bold).}
\label{bush}
\end{figure}
 
The solutions of the Gauss law (\ref{mcs-gauss}) for states in higher
Landau levels are $\Phi$ polynomials involving both $\ov B$ and $\ov{A}$
matrices: given that they transform in the same way under the gauge group, 
these polynomials can be represented by
bushes whose stems are  arbitrary words of $\ov{B}$ and $\ov{A}$,
as shown in Fig.(2b, 2c), where
$\ov{A}_{ab}$ is represented by a bold segment.
Since two stems cannot be equal, one obtains a kind of Fermi sea 
of $N$ ``one-particle states'' corresponding to the $N$ strands.
However,  there are additional degeneracies with 
respect to an ordinary fermionic system, because in each stem all possible
words of $\ov{A}$ and $\ov{B}$ of given length yield independent states
(for large $N$), owing to matrix noncommutativity, 
as seen in fig.2(b) and 2(c).

The complete filling of all the available degenerate $E>0$ states 
clearly gives very dense and inhomogeneous fluids that are incompatible
with the physics of the quantum Hall effect. Moreover, the matrix degeneracies
lead to a  density
of states in the $g=0$ theory that grows exponentially with the energy.
This is a characteristic of string theories that is
not suitable for the Hall effect \cite{mtheory}.
Of course, for $g>0$ the potential $\Tr [X,\ov{X}]^2$ restricts
the matrix noncommutativity and reduces the degeneracy: 
at $g=\infty$, this is not present and the theory can describe
a physical electron system, as is clear from the discussion of section 2.2.  

Given that the $g>0$ theory is very difficult to solve, in ref.   
\cite{c-rod} we introduced  a set of projections that
limit the matrix degeneracy at $g=0$ and are explicitly solvable.
These projections are expressed by the following 
constraints on the wave function,

\be
\left(A_{ab}\right)^m\ \Psi\ = \ 0\ \longrightarrow\ 
\left(\frac{\de\ \ }{\de \ov{A}_{ab}}\right)^m
\ \Phi(\ov{A},\ov{B},\psi) \ =\ 0\ ,\qquad
\forall\ a,b \ ,
\label{sll-cond}
\ee
for a given value of $m$. The $m=1$ case is the lowest Landau level
discussed before with no $\ov{A}$ dependence, while $m$ taking 
successive values $m=2,3,\dots$ gradually allow larger $\ov{A}$ 
multiplicities and thus matrix degeneracies.
Note that in equation (\ref{sll-cond}), each matrix component 
$A_{ab}$ is raised to the $m$-th power,  without index summation:
the condition is nevertheless gauge invariant and it admits an equivalent
manifestly invariant form that is discussed in section 3.

The results of ref.\cite{c-rod} were rather interesting: not only 
the projections (\ref{sll-cond}) allow homogeneous ground states suitable 
for describing quantum Hall fluids, but also they precisely occur in
the Jain pattern of filling fractions, $\nu=m/(mk+1)$, and their
derivation repeats step-by step the Jain ``composite fermion''
construction \cite{jain}.
 
Let us recall the main points of the analysis of \cite{c-rod}. 
Consider first the projection (\ref{sll-cond}) for $m=2$: 
the solutions are polynomials that are at most linear
in each component $\ov{A}_{ab}$.
Let us imagine that one or more $\ov{A}$ matrices are present at points on 
the bush as in Fig.(\ref{bush}).
The differential operator  (\ref{sll-cond})
acts by sequentially erasing pairs of bold lines on the bush,
each time detaching
two strands and leaving four free extrema with indices fixed to 
either $a$ or $b$, with no summation on them.
For example, when acting on a pair of $\ov{A}$ located on the same stem, 
it yields a non-vanishing result: this limits the bushes to have 
one $\ov{A}$ per stem at most.
The $A^2 \ap 0$ conditions can be satisfied if cancellations occur 
for pairs of $\ov{A}$ on different stems, as it follows:
\be
\left(A_a^b\right)^2\Phi =\cdots \ +\  \eps^{\dots i \dots j \dots}
\left(
\cdots M_{ia}\ N_{ja}\cdots V^b\ W^b \right)
\ +\ \cdots\ ,
\qquad (a,b\ {\rm fixed}).
\label{sll-canc}
\ee
This expression vanishes for $M=N$ due to the
antisymmetry of the epsilon tensor.
The general solution of (\ref{sll-cond}) is given by bushes involving one
$\ov{A}$ per stem at most (max N matrices in total), with all of
them located at the same height on the stems \cite{c-rod}. In formulas:
\ba
\Phi_{\{n_1,\dots,n_\ell; p; n_{\ell+1},\dots,n_M\}} 
&=& \eps^{i_1\dots i_N} \ 
\prod_{k=1}^{\ell}\ 
\left(\ov{B}^{n_k}\psi\right)_{i_k} \
 \prod_{k=\ell+1}^{N}\ 
\left(\ov{B}^p\ov{A}\ \ov{B}^{n_k} \psi\right)_{i_k}\ , 
\nl
&&\quad 0\le n_1 < \cdots < n_\ell\ ,\quad
0\le n_{\ell+1} < \cdots < n_N\ . 
\label{two-ll}
\ea

If the matrices $\ov{A},\ov{B}$ were diagonal, 
these states would be Slater determinants of ordinary Landau levels.
The matrix states have further degeneracies by commuting $\ov{A},\ov{B}$
pairs: however, commutations are almost
impossible in the the solution (\ref{two-ll}) of the $A^2 \ap 0$ constraint, 
were it not for the $p$ dependence. This shows how the
projection works in reducing matrix degeneracies.

The ground state in the $A^2 \ap 0$ theory 
with finite-box conditions corresponds to {\it homogeneous} filling 
all the allowed states in the first and second Landau levels with $N/2$
``gauge invariant particles'' each. It reads:
\be
\Phi_{1/2, \ gs} = \eps^{i_1\dots i_N} \ 
\prod_{k=1}^{N/2}\ 
\left(\ov{B}^{k-1}\psi\right)_{i_k} \ 
\prod_{k=1}^{N/2}\ 
\left(\ov{A}\ \ov{B}^{k-1} \psi\right)_{i_{N/2+k}}\ , 
\label{two-jain}
\ee
with angular momentum ${\cal J}=N(N-4)/4$.
This state is non-degenerate due to the vanishing of the 
$p$ parameter in (\ref{two-ll}). It has filling fraction
$\nu^*=2$, assuming homogeneity of its density\footnote{
This will be shown in section 4.}.

The ground states for $k>1$ are products of $k$
bushes: they obey the constraint $A^2\ap 0$ provided that 
the two derivatives always vanish when distributed over the bushes.
Given one bush of type (\ref{two-jain}),  
obeying $A^2\ \Phi_{1/2, \ gs}=0$, one can form the state,
\be
\Phi_{k+1/2, \ gs} \ = \ \Phi_{k-1, \ gs} \ \Phi_{1/2, \ gs} \ ,
\label{two-k-jain}
\ee
where the other $(k-1)$ bushes 
satisfy $A\ \Phi_{k-1,\ gs} =0$ and actually are Laughlin's one
(\ref{HVR-states}).
The angular momentum value for this state corresponds to the
filling fraction $1/\nu=k+1/2$ (cf (\ref{nu-def})).

We thus find the important result that the $A^2\ap 0$ projected
Maxwell-Chern-Simons theory possesses non-degenerate ground states 
that are the matrix analogues of the Jain states  obtained by 
composite-fermion transformation at $\nu^*=2$, 
i.e. $1/\nu=1/\nu^* + k$.
The matrix states (\ref{two-k-jain},\ref{two-jain}) 
would actually be exactly equal to Jain's wave functions, if the
$\ov{A},\ov{B}$ matrices were diagonal: the $\psi$ dependence
would factorize and the matrix states 
reduce to the Slater determinants of Jain's wave functions
(before their projection to the lowest Landau level) 
\cite{jain}\cite{numeric}.

The correspondence extends to the whole Jain series:
the other $\nu^*=m$ non-degenerate ground states, corresponding to
$\nu=m/(mk+1)$,
 are respectively obtained in the theories with $A^m\ap 0$ projections.
They read:
\be
\Phi_{k+1/m, \ gs}\ =\  \Phi_{k-1, \ gs} \ \Phi_{1/m, \ gs} \ ,
\label{m-k-jain}
\ee
where,
\be
\Phi_{1/m, \ gs} =  \eps^{i_1\dots i_N} \ 
\prod_{k=1}^{N/m}\ \left[
\left(\ov{B}^{k-1}\psi\right)_{i_k} \ 
\left(\ov{A}\ \ov{B}^{k-1} \psi\right)_{i_{k+N/m}}
\cdots
\left(\ov{A}^{m-1} \ \ov{B}^{k-1} \psi\right)_{i_{k+(m-1)N/m}}
\right]\ .
\label{m-jain}
\ee
Note that in the $A^m \ap 0$ theory, the
lower density states that were non-degenerate in the 
$A^k \ap 0$ theories, $k<m$, become degenerate; actually, 
when the boundary potential is tuned for letting (\ref{m-jain})
to be the ground state, the former states $(k<m)$
become excited states.
In conclusion, in ref.\cite{c-rod} we found that the ground states 
of the properly projected Maxwell-Chern-Simons matrix theory 
reproduce the Jain pattern of the composite fermion construction \cite{jain};
the matrix states are non-degenerate for specific values of the
density that are controlled by the boundary potential.

These results indicate that the Jain composite fermions
have some relations with the D0-brane degrees of freedom
and their underlying gauge invariance. 
Both of them have been described as dipoles: according to Jain
\cite{jain} and Haldane-Pasquier \cite{pasquier}, 
the composite fermion can be considered as the bound state
of an electron and a hole (a vortex in the electron fluid).
On the other side, matrix gauge theories, and the equivalent noncommutative
theories \cite{mtheory}, describe D0 branes that are point-like dipoles 
in the low-energy limit of string theory.
Although further relations between these two dipole pictures remain to
be found, in this paper we shall find further evidences that
the matrix ground states describe semiclassical incompressible fluids
with some of the properties of Hall states.

\begin{table}
\begin{center}
\[
\begin{array}{c|lccc|c|l}
m  & \# p_i=1 & \# p_i=2 & \# p_i=3 & \# p_i=4 & 1/\nu=1+1/\nu_{cl} & E/\B \\
\hline 
1  & k     &      &       &      &  k+1      &  0  \\
\hline 
2  & k-1   & 1    &       &      &  k+ 1/2   &  N/2   \\
\hline 
3  & k-2   & 2    &       &      &  k        &  N   \\
3  & k-1   & 0    & 1     &      &  k+1/3    &  N  \\
\hline 
4  & k-3   & 3    &       &      &  k-1/2    &  3N/2  \\
4  & k-2   & 1    & 1     &      &  k-1/6    & 3N/2   \\
4  & k-1   & 0    & 0     & 1    &  k+1/4    & 3N/2   
\end{array}
\]
\end{center}
\caption{Examples of standard (\ref{m-k-jain}) and 
generalized  (\ref{ext-jain}) Jain states for fixed value of
$k$, ordered by the level $m$ of projection, $ A^m\approx 0$, and their
 filling fraction $\nu$ and energy $E$ (disregarding the confining potential). 
Note that the experimentally  relevant values are $k=2,4$. }
\label{jain-tab}
\end{table}


\subsection{Generalized Jain's hierarchical states}

In the $A^m\ap 0$ projected theories with $m\ge 3$, there are other
solutions of the Gauss law for $k>1$ besides the Jain states (\ref{m-k-jain}).
These are obtained by combining products of any $k=1$ solution (\ref{m-jain}),
as follows:
\ba
\Phi_{\frac{1}{\nu}, \ gs} 
&=& \prod_{i=1}^k\ \Phi_{\frac{1}{p_i}, \ gs}\ ,\qquad\qquad
\frac{1}{\nu} \ =\ 1\ + \ \sum_{i=1}^k\ \frac{1}{p_i}\ ,
\nl
A^q\ \Phi_{\frac{1}{\nu},\ gs} &=& 
0\ ,\qquad\qquad \qquad\qquad q=1+\sum_{i=1}^k (p_i-1) \ .
\label{ext-jain}
\ea
In this equation, we also wrote the associated filling fractions
using Eq.(\ref{nu-def}), i.e. assuming homogeneous densities, and
the condition $A^q \approx 0$ that they satisfy.
The Jain mapping to a single set of  $\nu^*=q$
effective Landau levels does not hold for these generalized states
(having more than one $p_i>1$).
Actually, analogous generalized states were considered by Jain 
in his composite-fermion theory \cite{jain}, but
were discarded due to their instability (small or vanishing numerical gap).

Let us compare the generalized (\ref{ext-jain}) and 
standard (\ref{m-k-jain}) Jain states at fixed values of the background $k$
(keeping in mind that the physical values are $k=2,4$). 
The energy of the generalized states is additive in the 
$\nu^*=p_i$, $k=1$, blocks and reads:
\be
E_{\frac{1}{p_1}+\cdots+\frac{1}{p_k}, \ gs} 
= \frac{\B N}{2} \ \sum_{i=1}^k\ (p_i-1)\ +\ V_C\ .
\label{en-gen}
\ee
The analysis of some examples shows that
these additional solutions have in general higher energies for
the same filling or are more compact for the same energy
than the standard Jain states (\ref{m-k-jain}) (see Table \ref{jain-tab}).
States of higher energies  are clearly irrelevant at low temperatures;
higher-density states strongly deviate from the
semiclassical incompressible fluid value $\nu=1/(k+1)$ for background 
$\B\th=k$, that is specific of the Laughlin factors \cite{susskind}.
The analysis of the corresponding semiclassical states in section 4 will show
that most of these states are not incompressible fluids
(density not piecewise constant).

\section{Properties of the projection $A^m \ap 0$}

In this section we discuss the physical meaning of the projection:
\be
\left(A_{ab}\right)^m\ \Psi\left(\ov{A},\ov{B} \right)\ = \ 0\ ,
\qquad \forall a,b,
\label{proj-m}
\ee
that limits the degeneracy of matrix quantum states at $g=0$.
Although the operator $(A_{ab})^m$ is not gauge invariant, its kernel
restricted to gauge invariant states yields
a gauge invariant condition\footnote{
A formal proof of this statement is given in Appendix A},
as explicitly seen in section 2.
Therefore, there should exist a manifestly gauge invariant expression
for this condition, that is found in this section.
 
A simple example is useful to clarify the following discussion. 
In a two dimensional quantum mechanical
problem with rotation invariance ($O(2)$ global symmetry),
we consider the condition:
\be
P_m\ \Phi\ \equiv\ 
\left(\frac{\de}{\de x}\right)^m \ \Phi\left( r^2 \right)\ =\ 0\ ,
\qquad r^2=x^2+y^2\ ,
\label{ex-cond}
\ee 
where $\Phi$ is a reduced (polynomial) wave function.
The condition is not $O(2)$ invariant but its kernel acting on rotation
invariant functions does: indeed, it limits the order of the polynomial 
to $O(r^{m-1})$.
This example suggests two remarks:
\begin{itemize}
\item
The condition (\ref{ex-cond})
can have many different forms, that correspond to 
points on its orbit in the ``gauge'' $O(2)$  group: for example, 
an equivalent form is 
$(\de/\de y)^m \Phi=0$, corresponding to a $\pi/2$ rotation.
All these conditions are equally satisfied.
\item
A manifestly gauge-invariant expression can be obtained
by integrating over the gauge orbit, as follows:
\be
P_m \ \longrightarrow \ P_m^{g.i.}\ =\ 
\int_0^{2\pi} d\th \ \left( \cos\th\ \frac{\de}{\de x}\ +\  
\sin\th\ \frac{\de}{\de y}\right)^m\ . 
\label{int-cond}
\ee
\end{itemize}
However, this vanish for $m$ odd: the average looses information
because the operator $(\de/\de x)^m$ is not positive definite. Clearly, 
it can be made positive (and gauge invariant) by
contracting with another gauge-dependent term to obtain powers of the 
dilatation operator $D^m=(x^i \de/\de x^i)^m$.

We are now going to follow analogous steps for the condition $A^m\ap 0$.
First we find an equivalent, more general form.
Consider an infinitesimal $SU(N)$ gauge transformation $U=1+i\eps T$:
the Hermitean matrix $T$ can be expressed by the matrices
$E^{(ij)}$ with a single non-vanishing component,
$E^{(ij)}_{ab}=\d^i_a\d^j_b$, in symmetric or antisymmetric combinations,
$T=E^{(ij)}+E^{(ji)}$ or $T= i( E^{(ij)}- E^{(ji)})$.
Upon performing the gauge transformation, the $m=2$ constraint 
(\ref{sll-cond}),
$(U^\dag AU)_{ab}^2$, acquires an additional $O(\eps)$ term
that should also vanish on the wave functions obeying, $(A_{ab})^2\Psi=0$:
\be
0\ \approx\ A_{ab}\ \left[E^{ij},A \right]_{ab} \ = \ 
A_{ab} \left(\d_{ai}\ A_{jb} -A_{ai}\ \d_{jb}\right)\ , \qquad
\forall\ i\neq j \ ,\quad \forall\ a,b.
\label{trans-cond}
\ee 
We now analyse the various cases: 
\begin{itemize}
\item 
I. If $a=b$ and $i=a$ or $j=a$, we obtain the conditions,
$$0\ \approx\ A_{aa}\ A_{ja}\ \ap \ A_{aa} \ A_{ai} ,
\qquad \forall\ i,j\neq a\ . $$
\item
II. If $a\neq b$, we obtain,
\ba
{\rm 1.\ for}\ \ i=a\ {\rm and}\ j\neq b 
&\  \longrightarrow\  &
0\ \approx\ A_{ab}\ A_{jb}\ ,\quad \forall\ j\neq a,b\ , \nl
{\rm 2.\  for}\ \ i\neq a\ {\rm and}\ j = b 
&\  \longrightarrow\  &
0\ \approx\ A_{ab}\ A_{ai}\ ,\quad \forall\ \ i\neq a,b\ , \nl
{\rm 3.\  for}\ \ i= a\ {\rm and}\ j= b 
&\  \longrightarrow\  &
0\ \approx\ A_{ab}\left( A_{bb} -A_{aa}\right)\ . \nonumber
\ea
Note that each term in the linear combination
of case II.3 independently vanishes by case I.
\end{itemize}
These conditions can be summarized as follows:
\ba
A_{ab}\ A_{a'b}\ \Psi &=& 0\ ,\qquad \forall\ a,a',b, 
\nl
A_{ab}\ A_{ab'}\ \Psi &=& 0\ ,\qquad \forall\ a,b,b'.
\label{gen-pro2}  
\ea
They are more general than the original expression (\ref{sll-cond}) for $m=2$,
corresponding to $a=a'$ or $b=b'$.
Of course, iteration to $O(\eps^2)$ of the gauge transformation
produce further identities: these involve linear
combinations of $A^2$ terms that are not particularly useful;
for example, one such condition is: 
$A_{ab} A_{jc} + A_{jb} A_{ac}\approx 0$.

The $O(\eps^2)$ analysis is necessary to obtain the generalized constraint 
for $m=3$: the $O(\eps)$ expression is similarly,
$A_{ab}\ A_{ab}\ A_{ab'}\approx 0$, and its further transformation yields,
$$
0\ \approx\ 2\ A_{ab}\ \left[E^{ij},A \right]_{ab} \ A_{ab'}\ +
\ A_{ab}\ A_{ab}\ \left[E^{ij},A \right]_{ab'} \ .
$$
This expression contains the $m=3$ constraint analogous to (\ref{gen-pro2}):
\ba
A_{ab}\ A_{a'b}\ A_{a''b}\ \Psi &=& 0\ ,\qquad \forall\ a,a',a'',b, 
\nl
A_{ab}\ A_{ab'}\ A_{ab''}\ \Psi &=& 0\ ,\qquad \forall\ a,b,b',b'',
\label{gen-pro3}  
\ea
together with other relations involving linear combinations of cubic terms.

Following the $O(2)$ example, we can now transform the new expressions
(\ref{gen-pro2}) into positive definite operators.
We recall that the lowest Landau level condition corresponds
to the vanishing of the total energy, that is a sum of positive terms:
\be
H\ =\ \Tr \left( A^\dag\ A \right) \ = \ 
\sum_{a,b}\ A^*_{ab}\ A_{ab}\ \approx\ 0
\qquad \Leftrightarrow\qquad
A_{ab}\ \approx\ 0\ ,\ \ \forall \ a,b\ .
\label{lll}
\ee
We can construct the following positive definite expressions:
\ba 
Q_2 &=& \sum_{a,b,b'}\ A^\dag_{b'a}\ A^\dag_{ba}\ A_{ab}\ A_{ab'}\ ,
\label{q2-cond}\\
Q_2' &=& \sum_{a,a',b}\ A^\dag_{ba'}\ A^\dag_{ba}\ A_{ab}\ A_{a'b}\ ,
\label{q2p-cond}
\ea
whose vanishing is equivalent to the $m=2$ conditions (\ref{gen-pro2}).
These quantities are not yet gauge invariant but are convenient for the 
physical interpretation. We introduce the (gauge variant) energy operators
for one-particle matrix states, that are summed
over matrix indices of one row or column of $A_{ab}$, $Z_a$ or $Z'_b$, 
respectively:
\be
Z_a\ =\ \sum_b\ A^\dag_{ba}\ A_{ab}\ ,\qquad
Z'_b\ =\ \sum_a\ A^\dag_{ba}\ A_{ab}\ .
\label{zetas}
\ee 
Using these energy operators, we can rewrite (\ref{q2-cond},\ref{q2p-cond})
as follows:
\be
Q_2\ =\ \sum_a\  Z_a\left(Z_a\ -\ 1\right)\ ,\qquad
Q_2'\ =\ \sum_b\  Z_b'\left(Z_b'\ -\ 1\right)\ .
\label{q2zed}
\ee
In this form, the constraints $Q_2\Psi=Q_2'\Psi=0$ admit the following physical
interpretation:
there is a gauge choice in which the allowed states contains at most one
``particle'' in the second Landau level (energy equal to one) for
$(a,b)$ indices belonging to each row and column.

The constraint for $m=3$ (\ref{sll-cond}) similarly becomes:
\be
Q_3\ =\ \sum_a\  Z_a\left(Z_a\ -\ 1\right)\ \left(Z_a\ -\ 2\right)\ ,\qquad
Q_3'\ =\ \sum_b\  Z_b'\left(Z_b'\ -\ 1\right)\ \left(Z_b'\ -\ 2\right)\ .
\label{q3zed}
\ee
This requires that there at most 2 particles in the second Landau level 
or a single particle in the third level for any set of indices 
in a row or column.
The matrix labels are not gauge invariant, then these occupancies 
are only verified in specific gauges; nevertheless, the present form
of the constraints can be implemented in the semiclassical limit on
expectation values, $\la A_{ab}\ra$, as explained in section 4.

Next, we obtain the gauge-invariant form of the constraint $Q_2,Q_2'$
by averaging over the gauge group. We define:
\be
Q_2^{g.i.}\ = \ \int {\cal D} U\ Q_2'(U)\ =\ 
\sum_b \ \int {\cal D} U\ U^\dag_{bi}\ A^\dag_{ia'}\ U^\dag_{bj}\
A^\dag_{ja} \ A_{ak}\ U_{kb}\ A_{a'l}\ U_{lb}\ , 
\label{q2-int}
\ee
where ${\cal D}U$ is the invariant Haar measure \cite{samuel}.
The integrand is positive definite for any $U$ value,
because it can be thought of as the norm of a vector:
$Q_2(U)\sim \sum_b\ \vert A\cdot v^{(b)} \vert^4$, where
$v^{(b)}_a=U_{an}\d_n^b$ is a rotated unit vector. 
Therefore, we do not loose any information by performing the group average.

Group integrals of products of $U,U^\dag$ matrices can be found
e.g. in ref.\cite{samuel}: their results can be described as follows.
Representing the unitary matrices with upper and lower indices,
$U_{ab}\to U_a^{\ b}$, $(U^\dag)_{ab}\to (U^\dag)^a_{\ b}$, the result
of integrating $n$ $(U,U^\dag)$ pairs is a combination of $n$ delta functions
relating the upper indices among themselves times other $n$ 
deltas connecting the lower indices. The simplest integral is:
$$
\int{\cal D}U\ (U^\dag)^a_{\ a'}\ U_{b'}^{\ \ b}\ =\ 
\frac{1}{N}\ \d^{ab}\ \d_{a'b'}\ .
$$
In the general case of $n$ $(U,U^\dag)$ pairs, the pairings of upper (lower) 
indices by delta functions follow patterns given by the permutation 
of $n$ elements, with specific weight for 
each conjugacy class of permutations \cite{samuel}. For $n=2$, one finds:
\ba
\!\!\!\! && \int{\cal D}U\ (U^\dag)^a_{\ a'}\ U_{b'}^{\ \ b}\ 
(U^\dag)^c_{\ c'}\ U_{d'}^{\ \ d}
\ =\ \nl
\!\!\!\!\! && \ \ \frac{1}{N^2-1}\left[ 
\d^{ab} \d^{cd} \d_{a'b'} \d_{c'd'}\ +\ 
\d^{ad} \d^{cb} \d_{a'd'} \d_{c'b'} - \frac{1}{N}\left(
\d^{ab} \d^{cd} \d_{a'd'} \d_{c'b'}\ +\ 
\d^{ad} \d^{cb} \d_{a'b'} \d_{c'd'}\right)\right].\nonumber
\ea
In the case of the constraint $Q_2'$  (\ref{q2p-cond}),
 all the upper indices are simultaneously 
taking the same value $b$; thus, the different delta-function pairings
of upper indices take the same unit value.
As a result, the pairings of lower indices get averaged over,
and reduce to a plain sum over all pair permutations:
\be
Q_2^{g.i.}\ \propto \ \left(\d_{ki}\ \d_{lj}\ +\ \d_{kj}\ \d_{li}\right)\ 
A^\dag_{ia'}\ A^\dag_{ja} \ A_{ak}\ A_{a'l}\ . 
\label{q2-gi}
\ee
Upon commuting the operators to bring summed indices close each other, we
finally find the manifestly gauge-invariant form of the $A^2\ap 0$ constraint
(disregarding the normalization):
\be
Q_2^{g.i.}\ap 0\ , \qquad\quad
Q_2^{g.i.}\ = \ \Tr\left(A^\dag A A^\dag A\right)\ 
+\ \left( \Tr\ A^\dag A \right)^2\ -\ (N+1)\ \Tr\left( A^\dag A\right)\ .
\label{q2-gi-fin}
\ee
The same expression is also obtained by group averaging the other
operator $Q_2$ in (\ref{q2-cond}).
One can check that the action of the gauge-invariant constraint 
$Q^{g.i.}_2$ on bush wave functions (cf. section 2) is completely equivalent 
to that of the gauge-variant condition $A^2\ap 0$ \cite{c-rod}.

The gauge invariant form of the $m=3$ constraint can be similarly obtained 
by group averaging (\ref{gen-pro3}), leading to: 
\be
Q_3^{g.i.}\ =\ \sum_{\s\in{\cal S}_3}\
A^\dag_{i_1b}\ A^\dag_{i_2b'}\  A^\dag_{i_3b''}\ 
A_{i_{\s (3)}b''}\ A_{i_{\s (2)}b'}\ A_{i_{\s (1)}b}\ .
\label{q3-gi}
\ee
The form of this expression corresponding to (\ref{q2-gi-fin}) 
is not particularly illuminating.
The gauge-invariant expression (\ref{q3-gi}) straightforwardly generalizes 
to higher $m$ values.

In conclusion, in this section we have found equivalent forms of the
projections $A^m\ap 0$ of $g=0$ matrix states: 
the first expression (\ref{q2zed},\ref{q3zed}) in terms of 
occupation numbers is useful for the semiclassical limit considered in
the next section; the second expression (\ref{q2-gi-fin},\ref{q3-gi}) 
is manifestly gauge invariant. In the latter form, the constraint 
can be added to the Hamiltonian with a large positive coupling constant
to realize a softer form of projection, where matrix states violating the 
constraint are now allowed but possess very high energy.
For example, the quasi-particles excitations over the 
Jain ground states $\nu=m/(mk+1)$ would be possible, while they are
absent in the $A^m\approx 0$-projected theory, as explained in section 3.
A detailed analysis of this issue is postponed to a following publication.


\section{Droplet ground state solutions}

In this section we study the $g=0$ Maxwell-Chern-Simons theory 
in the semiclassical limit: we solve the classical equation of motion
including the quantum constraints, first for the ground states and then for
the quasi-hole excited states. 
We shall find the semiclassical states that correspond to the quantum states
with homogeneous filling and 
composite-fermion structure of section 2 \cite{c-rod}. 
The motivations for this semiclassical analysis are twofold: on one side,
previous experience \cite{poly1}\cite{mtheory}\cite{hansson}\cite{hansson2} 
\cite{lambert} with noncommutative field theory
has shown that the classical fluid dynamics incorporates some properties
of the full quantum theory. From another side, it is know that the 
Laughlin states in the quantum Hall effect are incompressible fluids
that become semiclassical in the thermodynamic limit $N\to\infty$ 
\cite{ctz-class}\cite{sakita}.
The semiclassical ground states we find in this section are also incompressible
fluids which, we believe, may give rather accurate descriptions of
the quantum matrix states for large $N$ values \cite{wiegmann}.

Let us start by writing the classical equations of motion:
the Hamiltonian of the Maxwell-Chern-Simons theory at $g=0$ 
can be written as follows:
\ba
 H &=& 2\ \Tr \left( \ov{A} A \right)\ +\ 
\w\ \Tr \left(\ov{B} B \right)\ +\
 \Tr \left[ \Lambda \left( [\ov{A}, A ] + [ \ov{B} , B ] - k 
+ \psi\otimes \ov{\psi} \right) \right] \nl
 && +\ \sum_a\ \Gamma_a \left(Z_a -\g \right)\ 
+\ \sum_b\ \G'_b \left(Z'_a -\g' \right)\ , 
\qquad\qquad\qquad \g,\g'=0,1,\dots,m-1\ .
\label{ham-cl}
\ea
We set $\B=2$, $\B\th =k\in \Z$,  and included the Gauss law constraint
via the Hermitean Lagrange multiplier $\L$. The projection 
$A^m\approx 0$ analyzed in section 3 is enforced by adding
two other Lagrange multipliers $\G_a,\G'_b$ times the energies,
$Z_a\ =\ \sum_b\ \ov{A}_{ba}\ A_{ab}$,
$Z'_b\ =\ \sum_a\ \ov{A}_{ba}\ A_{ab}$,
of single-particle states with matrix indices summed over rows or columns.
We replace the nonlinear constraints (\ref{q2zed},\ref{q3zed}) 
with linear expressions 
involving the parameters $\g,\g'$ taking the allowed values of $Z_a,Z'_b$. 
Since the constraints are not gauge invariant, we shall 
assume that we work in a gauge where they take integer values.
The gauge-invariant form of the constraint (\ref{q2-gi-fin}) found 
at the end of sections 3 is also not convenient because it would lead to
non-linear equations of motion that cannot be solved analytically. 
For the same reason, we limit the
confining potential (\ref{conf-pot}) to the quadratic term: later we shall see
how to avoid ground state degeneracies that may arise with 
this potential.

We vary the Hamiltonian with respect to $\ov{A},\ov{B}$, canonically
equivalent to the original $X,\P$, and obtain the equations:
\ba
i\ \dot{A}_{ab} &=& 2 A_{ab}\ -\ \left[ \Lambda , A \right]_{ab}\ 
+\ A_{ab}\left( \G_a \ +\ \G'_b \right)\ ,
\label{eq-a}\\ 
i\ \dot{B} &=& - \left[ \Lambda, B \right]\ + \w \ B \ ,
 \label{eq-b}\\
G &=& \left[ \ \ov{A} , A \ \right]\ +\ 
\left[ \ \ov{B} , B \ \right]\ - k + \psi\otimes \ov{\psi} = 0 \ ,
\label{eq-g}\\
Z_a &=& \sum_b\ \ov{A}_{ba}\ A_{ab}\ =\ \g\ , 
\qquad\qquad\qquad\qquad \g=0,1,\dots, m-1\ .
\label{eq-z}\\
Z'_b &=& \sum_a\ \ov{A}_{ba}\ A_{ab}\ =\ \g'\ , 
\qquad\qquad\qquad\qquad \g'=0,1,\dots, m-1\ .
\label{eq-zp}
\ea
We first discuss ground state solutions corresponding to
$\dot{A}=\dot{B}=0$.


\subsection{Laughlin ground states}

We can gain some intuition on the matrix equations 
(\ref{eq-a},\ref{eq-b},\ref{eq-g},\ref{eq-z})
by recalling the solutions of the Chern-Simons matrix model \cite{poly1},
corresponding to the lowest Landau level in our theory.
In the case $A=\ov{A}= 0$, we have $B=\ov{X}$ and
the ground state equations reduce to:
\ba
\left[ \Lambda, B \right]\ =\ \w \ B \ , \qquad\qquad
G = \ \left[ \ \ov{B} , B \ \right]\ - k + \psi\otimes \ov{\psi}\ =\ 0 \ .
\ea
These are the commutation relations of a truncated quantum harmonic
oscillator, with $\Lambda$ playing the role of  Hamiltonian. 
We can diagonalize it by gauge choice and write standard oscillator
matrices in the ``energy'' basis, $|0\ra,|1\ra,\dots,|N-1\ra$
\cite{poly1}:
\ba
B &=& \sqrt {k} \ \sum_{n=1}^{N-1} \sqrt{n} \mid n \ra \la n-1 \mid \  , 
\qquad \ \L  = \w \ \sum_{n=1}^{N-1} n \mid n \ra \la n \mid \ \ ,
\nl
\psi &=& \sqrt{kN} \ \mid N-1 \ra \ . \
\label{poly-sol}
\ea
In matrix form for $N=3$:
\be
B=\sqrt{k}\left(\begin{array}{ccc} 
0 & 0 & 0  \\ 1 & 0 &  0  \\ 0 & \sqrt{2} & 0  
\end{array}\right)\ , 
\qquad \Lambda= \w \left(\begin{array}{ccc}   
1 & 0 & 0  \\ 0 & 2 & 0  \\ 0 & 0 & 3  
\end{array}\right)\ , 
\qquad \psi=\sqrt{kN}\left( \begin{array}{c} 
0 \\ 0 \\ 1 \end{array} \right) \ .
\label{poly-mat}
\ee
The solution is characterized by the angular momentum,
$J=\Tr\ \ov{B}\ B=k\ N(N-1)/2$ and by vanishing energy.

A good definition of the density of semiclassical fluids in matrix models
\cite{poly1} is given in terms of the gauge invariant eigenvalues 
of $R^2 = \ov{X}X $,
\be
\r(r^2)\ =\ \sum_{i=0}^{N-1}\ \d\left(r^2-\s_i \right)\ ,
\qquad\qquad \s_i \ \in {\rm Spec} \left(R^2 \right)\ .
\label{rho-def}
\ee
In the limit $N\to\infty$, this becomes a piecewise continuous function
that describes two-dimensional rotation-invariant distributions
($\r (r)=\r(r^2)/\pi$).
A discrete approximation is:
\be
\r(r^2)\ =\ \sum_{i}\ \frac{n_i}{\s_{i+1}-\s_i}\ 
\d_{r^2,\s_i}\ ,
\label{rho-disc}
\ee
involving the Kronecker delta and the ordered set of distinct 
$\s_i$ eigenvalues, $\s_i<\s_j$, $i<j$, with multiplicities $n_i$.

\begin{figure}
\begin{center}
\includegraphics*[width=8cm]{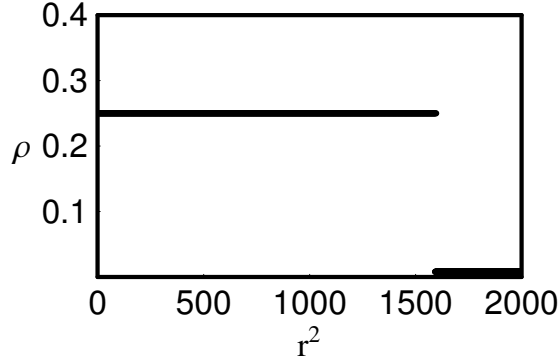}
\end{center}
\caption{Plot of the density for the semiclassical Laughlin ground state
(\ref{poly-sol}) for $\nu_{cl}=1/k$, $k=4$, and $N=400$.}
\label{fig3}
\end{figure}

In Polychronakos' solution of the Chern-Simons matrix model (\ref{poly-sol}),
the matrix $R^2$ is already diagonal,
\ba
R^2 \ = \ov{X} X \ = {\rm diag}\left(0,k,2k,\dots,(N-1)k \right)\ .
\label{poly-r2}
\ea
Its density (Fig. \ref{fig3}) is constant and describes a circular droplet of 
fluid with Laughlin filling fraction \cite{susskind}:
\ba
\nu_{cl} =\frac{2 \pi \rho_o}{e \textbf{B}}=\frac{1}{k} \ , 
\qquad \mbox{with} \quad \rho_o= \frac{1}{2\pi\theta} \ .
\label{Laugh-fill}
\ea
Note that an ordering ambiguity in the definition of $R^2$ was resolved
by matching to the angular momentum spectrum. As said before, there is a shift
in the filling fraction in quantum theory due to a 
contribution from the integration measure, $1/\nu =1+1/\nu_{cl}$; thus,
$k$ should be even for describing electrons.

\subsection{Jain ground states}

As we showed in section 2.3, the Maxwell Chern-Simons theory contains 
Jain-like ground states (\ref{m-k-jain}), 
that involve higher Landau levels ($A \neq 0$).
Their filling fractions can be written as in composite fermion 
construction \cite{jain}, 
\be
\frac{1}{\nu}\ =\ \frac{1}{\nu^*}\ + \ k\ +1 \ ,\qquad \quad k\ {\rm even},
\qquad\qquad \nu^*=2,3,\dots,
\label{jain-trans}
\ee 
and their energy and angular momentum values are recalled in Table 1. 
We first note that these states are characterized by
energies $O(N)$ and angular momentum $J=O(N^2)$, thus implying that
the matrix $A$ must have elements of $O(1)$ and be much
smaller than $B$.
Indeed, the constraints, $Z_a,Z'_b=0,1,\dots, m-1$, limits the
squares of $A_{ab}$ matrix elements  
summed over each row or column to take at most the total value $(m-1)$. 
Were it not for this constraint, the $A,B$ matrices could be
rescaled in the ground state equations (\ref{eq-a},\ref{eq-b},\ref{eq-g}) 
to eliminate the $k$ dependence, leading to solutions with $E=O(kN)$ at least.

We now describe the solution of the ground state
equations of motion in the $A^2\ap 0$ projected theory
($\g,\g'=0,1$ in (\ref{eq-z})).
Under some minor hypotheses, we find a single solution corresponding
to the unique quantum state with $\nu^*=2$ (\ref{two-k-jain}) \cite{c-rod}. 
Working in analogy with the Laughlin case
(\ref{poly-mat}), we shall try a distribution of $R^2$
eigenvalues leading to a piecewise constant density.
We can consider the gauge in which $\L$ is diagonal, $\L ={\rm diag}(\ell_a)$,
and assume that $\psi$ has a single non-vanishing component, i.e. the last
one, as in (\ref{poly-mat}), such that the term, 
$(k \ I\ -\psi\otimes\ov{\psi})$, is also diagonal.
The equation for $B$ (\ref{eq-b}),
\be
\left(\ell_a-\ell_b\right) B_{ab}=\w\ B_{ab}\ ,
\label{eq-b2}
\ee
requires that $B$ is a raising operator, i.e. non vanishing on a single
diagonal, $B_{ab}\propto\d_{a,b+n}$; moreover, the Gauss law (\ref{eq-g}) 
requires, $[\ov{B},B]\sim k\ I$, 
apart from $O(1)$ corrections due to $[\ov{A},A]$. Therefore, $B$ should be  
non-vanishing on the first diagonal:
\be
B_{ac}\ =\ \d_{a,c+1}\ b_{c+1}\ , \qquad\qquad c=0,\dots,N-2\ .
\label{b-sol}
\ee
Eq. (\ref{eq-b2}) implies evenly spaced $\L$ eigenvalues,
 $\ell_{a+1}-\ell_a=\w$, and leaves the components $b_c$ undetermined.
The equation (\ref{eq-a}) for $A$ reads:
\be
A_{ab}\ \neq\ 0\qquad\longrightarrow\qquad \G_a\ +\ \G'_b=(a-b)\w\ -\ 2 \ ,
\label{a-cond}
\ee
that can always be solved for $\G_a,\G'_b$.
The constraints, $Z_a,Z'_b=0,1$, imply that $A_{ab}$ has one 
non-vanishing element per row and column, at most, equal to one.
If it had exactly one element per row and column, it would be 
the representation of a permutation, $\s \in {\cal S}_N$. 
Therefore, we can write:
\be
A_{ab}\ = \ \d_{a,\s(b)}\ a_{b+1}\ ,\qquad \qquad a_b=0,1\ ,\qquad
\quad \s\ \in\ {\cal S}_N\ .
\label{a-sol}
\ee
We now consider the Gauss law (\ref{eq-g}): all terms in this equation
are diagonal matrices, leading to a system of $(N-1)$ scalar equations for the
$A,B$ matrix elements $\{a_b,b_b \}$.
Note that both matrices $\ov{A} A$ and $\ov{B} B$ are diagonal and thus their
elements are positive integers in the semiclassical theory:
$b_b^2\in \Z_+$.
After introducing,
\be
\b_b =b_b^2\ , \qquad\qquad\qquad \a_b=a_b^2\ ,
\label{be-al}
\ee
we obtain the system:
\ba
\b_1 &=& k - \a_1 + \a_{\s(1)}\ ,\nl
\b_2 - \b_1 &=& k - \a_2 + \a_{\s(2)}\ , \nl
\dots &=&\dots \dots\ ,\nl
\b_{N-1} -\b_{N-2} &=& k- \a_{N-1} + \a_{\s(N-1)}\ .
\label{gauss-sys}
\ea
The solution can be found by thinking to the expected shape
of the droplet. The quantum state (\ref{two-k-jain}) is made of $k$
generalized Slater determinants with homogeneous filling of $N$ 
``one-particle'' states\footnote{
This Fock-space analogy is meaningful for diagonal matrices, 
and may not be correct
in general: its limitations will be discussed in section 4.3.}.  
Each one-particle state is expected to give a constant contribution to the 
density of the droplet: there are $(k-1)$ Laughlin terms and one 
term with $N/2$ ``particles'' in the second Landau levels 
spanning half of the angular momentum range, as confirmed by the 
quantum numbers, $E=\B N/2$ and $J=(k-1+1/2)\ N^2/2+O(N)$.
The contribution are additive in terms of angular momentum eigenvalues,
$J\sim \Tr\ \ov{B}B=\sum_{i=1}^{N-1}\ \b_i$
(the $O(N)$ contribution of $\Tr\ \ov{A}A$ is subdominant for $N\to\infty$).
Therefore, we expect,
$\b_i  \sim  (k-1) i\ $, for one half of the range, say $0<i<N/2$,
and $\b_i  \sim  (k+1) i\ $ for the other half.
Moreover, $\b_i$ should be continuous at $i=N/2$ in order to obey
the corresponding equation with $\a_i=O(1)$.
We take:
\ba
\b_i &= & (k-1)\ i\ ,\qquad \qquad\qquad\qquad 0<i\le\frac{N}{2}\ ,\nl
\b_i &= & (k+1)\ i\ -\ N,\qquad \qquad\qquad \frac{N}{2}<i<N\ .
\label{beta-sol}
\ea
The solution for $A$ is found by inspection: it has $N/2$ non-vanishing
 elements equal to one on the diagonal of the lower half sector.

Summarizing, the ansatz semiclassical ground state solution for $\nu^*=2$
is given by ($N$ even):
\ba
B &=& \sum_{n=1}^{N/2} \sqrt{n(k-1)} \mid n \ra \la n-1 \mid\ + \  
\sum_{n=\frac{N}{2}+1}^{N-1} \sqrt{ n(k+1)-N} \mid n \ra \la n-1 \mid \ ,
\nl
A &=& \sum_{n=0}^{\frac{N}{2}-1} \mid n+ \frac{N}{2} \ra \la n \mid \ .
\label{nu2-sol}
\ea
In matrix form for $N=4$, it reads:
\be
B = \left(\begin{array}{cccc} 
0 & 0 & 0 & 0 \\ \sqrt{k-1} & 0 &  0 & 0 \\ 
0 & \sqrt{2(k-1)} & 0 & 0 \\ 0 & 0 & \sqrt{3k-1} & 0 
\end{array}\right)\ \ , \qquad
A = \left(\begin{array}{cccc}  
0 & 0 &  0 & 0 \\ 0 & 0 & 0 & 0 \\ 1 & 0 & 0 & 0 \\ 
0 & 1 & 0 & 0 \end{array}\right) . \
\label{nu2-mat}
\ee
This solution has same energy $E=\B N/2$ of the quantum state
(\ref{two-k-jain}) and same angular momentum $J=(k-1/2)N^2/2+O(N)$  to leading 
 order (cf. Table 1).
The matrix $R^2=(B+\ov{A})(\ov{B}+A)$ contains off-diagonal terms
from the mixed products: however, these give subdominant $O(1/\sqrt{N})$
corrections to the eigenvalues as is clear in a simple two-by-two matrix
example. Thus, ${\rm Spec}(R^2)= {\rm Spec}(B\ov{B})(1+O(1/\sqrt{N}))$,
confirming the earlier identification of droplet shape with angular momentum
spectrum.

\begin{figure}
\begin{center}
\includegraphics[width=9cm]{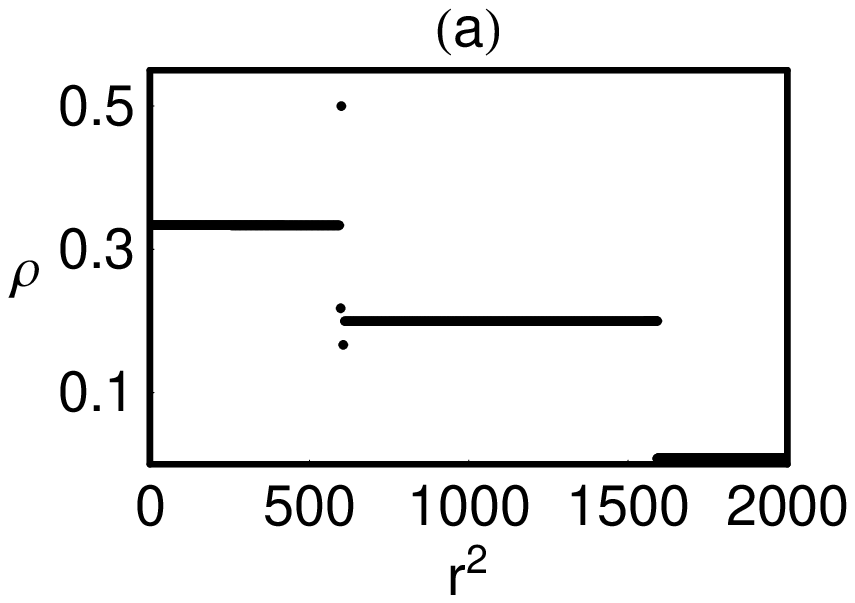}

\includegraphics[width=9cm]{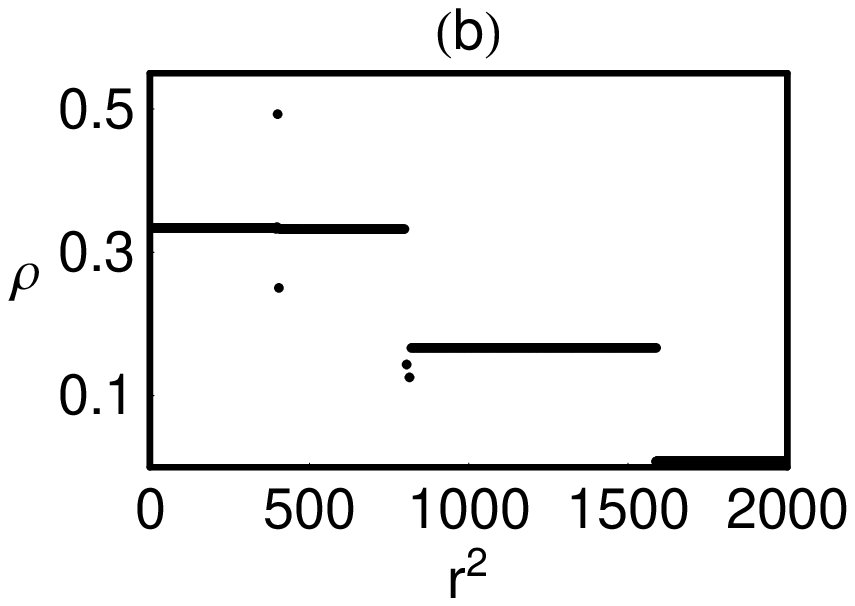}

\includegraphics[width=9cm]{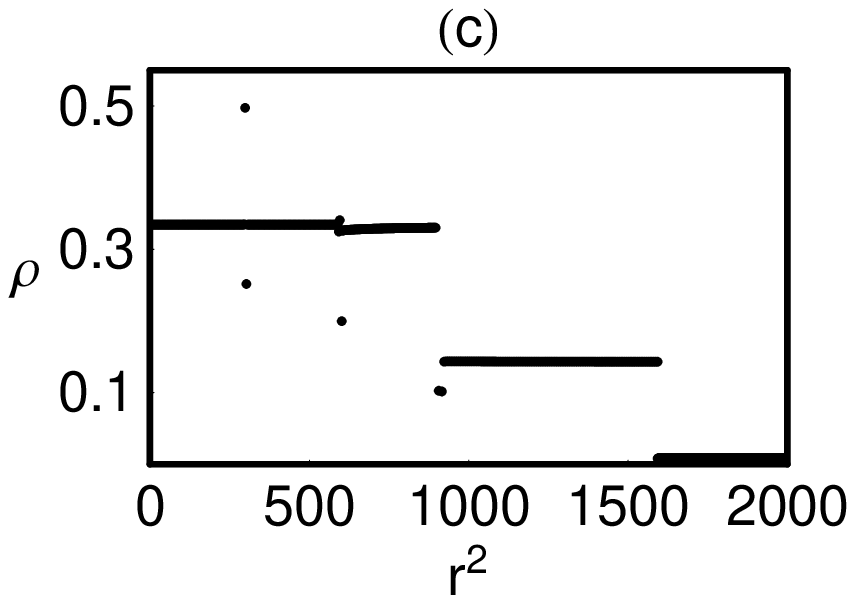}
\end{center}
\caption{Plot of the density for the Jain matrix ground states
with $1/\nu_{cl}=1/\nu^*+k$, for $k=4$ and $N=400$: 
(a) $\nu^*=2$ (\ref{nu2-sol});
(b) $\nu^*=3$ (\ref{nu3-sol}); and (c) $\nu^*=4$ (\ref{nu4-sol}).}
\end{figure}

In Fig.4(a), the density (\ref{rho-disc}) of the droplet of fluid is
plotted by computing the exact spectrum for $N=400$: 
up to finite-$N$ fluctuations,
this is a two-step constant density as anticipated.
We recall that the same droplet shape is found for
the Jain phenomenological states before their
projection to the lowest Landau level \cite{jain}; the density becomes
constant only after projection\footnote{ 
The lowest-level projection in the matrix theory cannot be done at
present, lacking an understanding of the $g>0$ regime: 
at $g=0$, it would give a trivial result because the Laughlin state is the 
unique lowest-level ground state for any $k$ value.}.

In ref.\cite{c-rod}, we argued that the matrix ground states at $g=0$ 
match one-to-one the phenomenological Jain states that are
good ansatz in the physical limit $g=\infty$:
the two sets of states become identical in the limit of both $\ov{X},X$ 
diagonal, that can be formally reached at $g=\infty$. 
To establish a relation at the quantum level, 
we would need to consider the evolution of the matrix ground states as the
coupling is varied in between, $0<g<\infty$, and to check that the gap never
vanishes, i.e. that there are no phase transitions in $(B,g)$ plane
(cf. Fig.1) separating the $g=0$ and $g=\infty$ regions
at these density (i.e. total angular momentum) values \cite{c-rod}.
While this behaviour remains to be proved, it is supported by the result that
the matrix ($g=0$) and phenomenological ($g=\infty$) states 
have similar densities of incompressible fluids.

We also note that the solution (\ref{nu2-sol}) could also be obtained in the 
lowest-level theory (Chern-Simons matrix model) by replacing
the $A$ matrix with $N/2$ different ``boundary'' auxiliary fields 
$\psi \to\psi_\a$, $\a=1,\dots,N/2$. This multi-boundary generalization
of Polychronakos' model has been considered in ref.\cite{poly3}: 
it naturally describes multicomponent droplets, i.e. 
$1/\nu=n/k$ for $n$ boundary fields. However, the description of Jain states
is rather unnatural, because the number of auxiliary fields is
macroscopic and should be adjusted
for each Jain state; moreover, this theory does not admit
the physical limit of commuting matrices.

The solution (\ref{nu2-sol}) can be easily generalized for the theory with
projection $A^3\approx 0$, possessing a Jain ground state with $\nu^*=3$:
this is found at the specific density that is reached
by tuning the boundary potential.
The constraint now allows the $A_{ab}$ components (\ref{a-sol}) 
to take values $a_b=0,1,\sqrt{2}$; 
we assume again a single non-vanishing element per row and column,
eq. (\ref{a-sol}), otherwise the commutator, $[\ov{A},A]$, 
would have off-diagonal terms that cannot be matched in the Gauss law
equation (\ref{eq-g}). 
Therefore, the equations (\ref{gauss-sys}) are unchanged.
Let us recall that the quantum solution contains 
$(k-1)$ Laughlin terms and the
$\nu^*=3$ piece that puts three particles in the same angular momentum
state, ranging from zero to $N/3$. Thus, the $B$ ansatz contains eigenvalues
spaced by $(k-1)$ for $2/3$ of the droplet and by $(k+2)$ for $1/3$ of it.
The matrix $A$ that solves the Gauss law (\ref{gauss-sys}) involves
elements on a diagonal extending for $2/3$ of the matrix ($N$ should be
a multiple of $3$). In conclusion:
\ba
B &=& \sum_{n=1}^{2N/3} \sqrt{n(k-1)} \mid n \ra \la n-1 \mid \ +\  
\sum_{n=\frac{2N}{3}+1}^{N-1} \sqrt{ n(k+2) -2N} \mid n \ra \la n-1 \mid\ ,
\nl
A &=& \sum_{n=0}^{\frac{N}{3}-1} \mid n+ \frac{N}{3} \ra \la n \mid \ +\  
\sum_{\frac{N}{3}}^{\frac{2N}{3}-1} \sqrt{2} \mid n \ +\ 
\frac{N}{3}\ra \la n \mid \ . \
\label{nu3-sol}
\ea
In matrix form for $N=6$:
\ba
B &=& \scriptsize{\left(\begin{array}{cccccc} 
0 & 0 & 0 & 0 & 0 & 0 \\ \sqrt{k-1} & 0 &  0 & 0 & 0 & 0 \\ 
0 & \sqrt{2(k-1)} & 0 & 0 & 0 & 0 \\ 0 & 0 & \sqrt{3(k-1)} & 0 & 0 & 0 \\ 
0 & 0 & 0 & \sqrt{4(k-1)} & 0 & 0 \\ 0 & 0 & 0 & 0 & \sqrt{5k - 2} & 0  
\end{array}\right)\ , }
\nl
A &=& \scriptsize{\left(\begin{array}{cccccc} 
0 & 0 & 0 & 0 & 0 & 0 \\ 0 & 0 &  0 & 0 & 0 & 0 \\ 1 & 0 & 0 & 0 & 0 & 0\\ 
0 & 1 & 0 & 0 & 0 & 0 \\ 0 & 0 & \sqrt{2} & 0 & 0 & 0 
\\ 0 & 0 & 0 & \sqrt{2} & 0 & 0 
\end{array}\right)\ } . \
\label{nu3-mat}
\ea
The droplet shape plotted in Fig.4(b) 
has again two steps, up to local fluctuations that vanish for $N\to\infty$.

The ansatz solution with $\nu^*=4$ in the theory $A^4\approx 0$ again
involves a matrix $B$ with two-speed spectrum and a matrix $A$
with elements $a_b=1,\sqrt{2},\sqrt{3}$, on the diagonal extending over
$3/4$ of the matrix ($N$ multiple of $4$):
\ba
B &=& \sum_{n=1}^{3N/4} \sqrt{n(k-1)} \mid n \ra \la n-1 \mid\ +\ 
\sum_{n=\frac{3N}{4}+1}^{N-1} \sqrt{n(k+3)-3N} \mid n \ra \la n-1 \mid\ ,
\nl
A &=& \sum_{n=0}^{\frac{N}{4}-1} \mid n+ \frac{N}{4} \ra \la n \mid \ + \ 
\sum_{\frac{N}{4}}^{\frac{2N}{4}-1} \sqrt{2} \mid n + 
\frac{N}{4}\ra \la n \mid \ +\  
\sum_{\frac{2N}{4}}^{\frac{3N}{4}-1} \sqrt{3} \mid n + 
\frac{N}{4} \ra \la n \mid \ .
\label{nu4-sol}
\ea
In matrix form for $N=8$:
\ba
B &=& {\sc \left(\begin{array}{cccccccc} 
0 & 0 & 0 & 0 & 0 & 0 & 0 & 0 \\ 
\sqrt{k-1} & 0 &  0 & 0 & 0 & 0 & 0 & 0 \\ 
0 & \sqrt{2(k-1)} & 0 & 0 & 0 & 0 & 0 & 0 \\ 
0 & 0 & \sqrt{3(k-1)} & 0 & 0 & 0 & 0 & 0 \\ 
0 & 0 & 0 & \sqrt{4(k-1)} & 0 & 0 & 0 & 0 \\ 
0 & 0 & 0 & 0 & \sqrt{5(k - 1)} & 0 & 0 & 0 \\ 
0 & 0 & 0 & 0 & 0 & \sqrt{6(k-1)} & 0 & 0 \\ 
0 & 0 & 0 & 0 & 0 & 0 & \sqrt{7k - 3} & 0 \end{array}\right)} \ ,
\nl
A &=& {\sc \left(\begin{array}{cccccccc} 
0 & 0 & 0 & 0 & 0 & 0 & 0 & 0 \\ 
0 & 0 &  0 & 0 & 0 & 0 & 0 & 0 \\ 
1 & 0 & 0 & 0 & 0 & 0 & 0 & 0 \\ 
0 & 1 & 0 & 0 & 0 & 0 & 0 & 0 \\ 
0 & 0 & \sqrt{2} & 0 & 0 & 0 & 0 & 0\\ 
0 & 0 & 0 & \sqrt{2} & 0 & 0 & 0 & 0 \\ 
0 & 0 & 0 &\mathbb{} 0 & \sqrt{3} & 0 & 0 & 0 \\ 
0 & 0 & 0 & 0 & 0 & \sqrt{3} & 0 & 0 \end{array}\right)}\ . 
\ea
The density for $N=400$ and $k=4$ is plotted in Fig.4(c).


\subsection{Correspondence of semiclassical and quantum states}

Here we provide a simple argument to support the identification 
of the semiclassical solutions with the quantum states of section 2.
Consider first the correspondence for the Laughlin states,
(\ref{Laugh-mat}) and (\ref{poly-sol}).
We choose the gauge in which the expectation values
of $B$ matrix elements on the quantum state 
take the classical values (\ref{poly-sol}) found in the previous
section, up to subleading corrections for $N\to\infty$.
Let us rewrite the $N=4$ wave function in terms of
these non-vanishing terms only\footnote{
Although this expansion should hold for $N\to\infty$, we write the 
$N=4$ case for simplicity; the expression for general $N$ 
can be easily inferred.}:
\ba
\Phi_{k,\ gs} &=& \left[
\eps^{a_1 a_2 a_3 a_4}\ \psi_{a_1}\ (\ov{B}\psi)_{a_2}
\ (\ov{B}^2\psi)_{a_3}
\ (\ov{B}^3\psi)_{a_4} \right]^k \nl
&\sim & \left[\eps^{3210}\ \psi_{3}\ (\ov{B}_{23}\psi_{3})
\ (\ov{B}_{12}\ov{B}_{23}\psi_{3})
\ (\ov{B}_{01}\ov{B}_{12}\ov{B}_{23} \psi_{3}) \right]^k \ .
\label{lau-class}
\ea
This ``semiclassical wave function'' describes ``particles'' with
matrix indices, $(01),(12),(23)$, in angular momentum states 
that precisely match the occupation numbers $\ov{B}_{ab}B_{ba}$ given
by the classical solution (\ref{poly-r2}), equal to $(k,2k,3k)$, 
respectively.  
This is a self-consistent argument for the correspondence of states:
in the semiclassical $N\to\infty$ limit, the quantum states match 
the semiclassical solutions for the leading occupation numbers.

A similar relation holds for the $\nu^*=2,3,4$ Jain states.
For $\nu^*=2$, the quantum wave function is (\ref{two-k-jain});
we evaluate it on the semiclassical non-vanishing 
$\ov{A}_{ab},\ov{B}_{ab}$ values (\ref{nu2-sol}), given explicitly for $N=4$:
\ba
\Phi_{k+1/2,\ gs} &=&
\left[\eps^{a_1 a_2 a_3 a_4}\ \psi_{a_1}\ (\ov{B}\psi)_{a_2}
\ (\ov{B}^2\psi)_{a_3}\ (\ov{B}^3\psi)_{a_4}\right]^{k-1} \nl
&&\times
\eps^{a_1 a_2 a_3 a_4}\ \psi_{a_1}\ (\ov{B}\psi)_{a_2}
\ (\ov{A}\psi)_{a_3}\ (\ov{A}\ov{B}\psi)_{a_4} \nl
&\sim & 
\left[\eps^{3210}\ \psi_{3}\ (\ov{B}_{23}\psi_{3})
\ (\ov{B}_{12}\ov{B}_{23}\psi_{3})
\ (\ov{B}_{01}\ov{B}_{12}\ov{B}_{23} \psi_{3}) \right]^{k-1} \nl
&&\times
\eps^{3210}\ \psi_{3}\ (\ov{B}_{23}\psi_{3})
\ (\ov{A}_{13}\psi_{3})
\ (\ov{A}_{02}\ov{B}_{23} \psi_{3}) \ .
\label{jain-class}
\ea
The ``one-particle'' occupancies of both energy and angular momentum states
given by the wave function again match
the expectation values of the corresponding
number operators,  $\ov{A}_{ab}A_{ba}$ and $\ov{B}_{ab}B_{ba}$, 
of the classical solution.
The correspondence extends to the other $\nu^*=m$ states that have spectrum
of occupancies given by (\ref{nu3-sol},\ref{nu4-sol}).
This argument support our belief that the large $N$ limit
of the matrix theory is semiclassical for the incompressible 
fluid ground states (piecewise constant density) and their small excitations.


\subsection{Generalized Jain states}

In the analysis of \cite{c-rod}, we found other quantum solutions to the
constraint $A^m\approx 0$, for $m\ge 3$, besides Jain composite fermion
wave functions. They were recalled in section 2.4, eq. (\ref{ext-jain})
and summarized in
Table 1: these are analogs of Jain's generalized hierarchical states,
made by products of two or more wave functions with higher-level
fillings ($p_i> 1$).
In the semiclassical analysis, we find that some of these states have
corresponding solutions with piecewise constant density, while most of
them do not. Besides, we find spurious ground states 
that are allowed by the simplistic quadratic boundary potential
used in (\ref{ham-cl}).
Let us describe these solutions in turn.

\subsubsection{Spurious solutions}

There is a variant of the  composite-fermion solution
for $m=3,4,\dots$, Eqs. (\ref{nu2-sol},\ref{nu3-sol}), 
where the $A$ matrix elements
take the same values, but their positions are permuted. 
For $m=3$, this is:
\ba
B &=& \sum_{n=1}^{N/3} \sqrt{n(k-2)} \mid n \ra \la n-1 \mid \ +\  
\sum_{n=\frac{N}{3}+1}^{N-1} \sqrt{ n(k+1) -N} \mid n \ra \la n-1 \mid\ ,
\nl
A &=& \sum_{n=0}^{\frac{N}{3}-1}\sqrt{2}  \mid n+ \frac{N}{3} \ra \la n \mid 
\ +\  \sum_{n=\frac{N}{3}}^{\frac{2N}{3}-1} \mid n \ +\ 
\frac{N}{3}\ra \la n \mid \ . \
\label{nu3-sol2}
\ea
The total energy and angular momentum values are the same as those of the 
$m=3$ Jain state, $E=\B N$, ${\cal J}\sim(k-2/3)N^2/2$ (cf. Table 1).
The corresponding $B$ matrix again describes a two-step droplet.
In order to find the corresponding quantum state, we use
the classical-quantum correspondence of the previous section.
The single-particle occupation numbers of the classical solution, 
for e.g. $N=6$, correspond to those of $(k-2)$ Laughlin factors 
plus the occupations 
$(3,6,9)$, $(2,2,1,1)$ and $(12)$ for the components 
$(\ov{B}_{23},\ov{B}_{34},\ov{B}_{45})$, 
$(\ov{A}_{02}\ov{A}_{13}\ov{A}_{24}\ov{A}_{35})$ and $\psi_5$, 
respectively.
These components should fit into two wave function of the type 
(\ref{two-ll}) that obey $A^2\ap 0$ (the product wave function obeying
$A^3\ap 0$). The solution, rewritten in gauge invariant form, is:
\ba
\Phi &=& \left(\Phi_{1,\ gs}\right)^{k-2}\ 
\left(\eps^{a_1 a_2 a_3 a_4 a_5 a_6}\ \psi_{a_1}\ (\ov{B}\psi)_{a_2}
\ (\ov{B}^2\psi)_{a_3}\ (\ov{B}^3\psi)_{a_4} 
\ (\ov{A}\ov{B}^2\psi)_{a_5}\ (\ov{A}\ov{B}^3\psi)_{a_6} \right)\nl
&&\times\ 
\left(\eps^{a_1 a_2 a_3 a_4 a_5 a_6}\ \psi_{a_1}\ (\ov{B}\psi)_{a_2}
\ (\ov{A}\psi)_{a_3}\ (\ov{A}\ov{B}\psi)_{a_4} 
\ (\ov{A}\ov{B}^2\psi)_{a_5}\ (\ov{A}\ov{B}^3\psi)_{a_6} \right).
\label{spur}
\ea
In this state, 
we recognize that some strands do not have minimal length: 
thus, this is an excited state for a Hamiltonian
with more realistic boundary terms (\ref{conf-pot}) realizing
finite-box conditions\footnote{
The quadratic potential used in (\ref{ham-cl}) is known to yield
such degeneracies (\cite{ctz-class}).}.
The expectation value of the higher boundary potential 
$\la \Tr\left( \ov{B}^2\ B^2\right)\ra$ 
on this state is actually larger than that of the 
$m=3$ Jain state (\ref{nu3-sol}) with same energy and angular momentum:
this confirms our interpretation of the solution (\ref{nu3-sol2}).

\subsubsection{Other two-step density states}

Among the generalized Jain state in Table 1, there are those made
by two kinds of terms, as follows:
\be
\Phi_{\frac{1}{\nu}, \ gs} \ = \
\left(\Phi_{1,\ gs}\right)^{k-n}\ 
\left(\Phi_{\frac{1}{2}, \ gs}\right)^{n} \ ,
\qquad\qquad \frac{1}{\nu} \ = \ \frac{n}{2}+ (k-n) +1\ ,
\qquad n=2,3,\dots\ ,
\label{2-halfs}
\ee
in the notation of Eq.(\ref{ext-jain}). They
obey, $A^{n+1}\approx 0$, for $n=2,3,\dots$, and violate
the composite-fermion transformation (\ref{jain-trans}) \cite{jain}.
In the droplet interpretation of classical solutions
of section 4.2, we expect a density of $R^2$ eigenvalues 
equal to $(k+n)$ for half of the spectrum and $(k-n)$ for the second half. 
The ansatz has the two-block structure of the solution (\ref{nu2-sol}),
with maximal value $A_{ab}=n$ in agreement with the constraint.

\begin{figure}
\includegraphics[width=8cm]{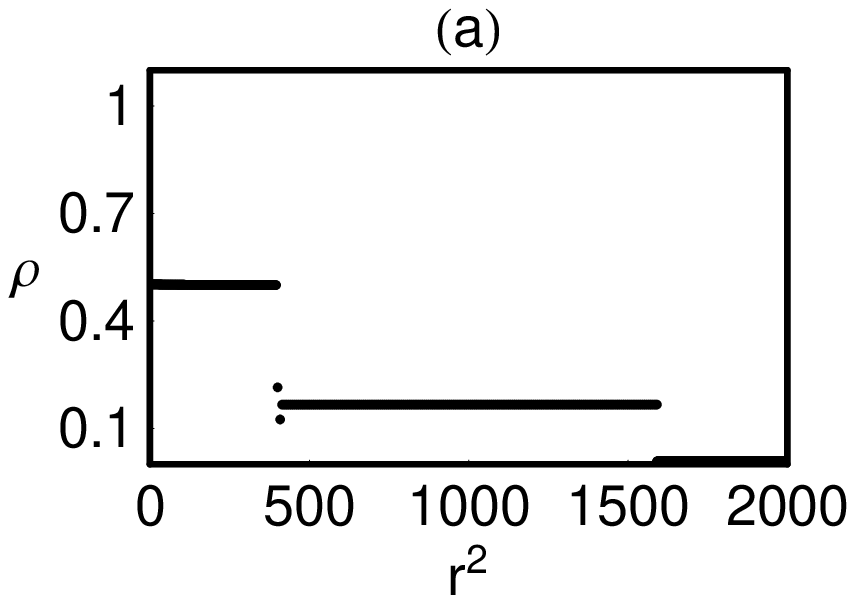}
\includegraphics[width=8cm]{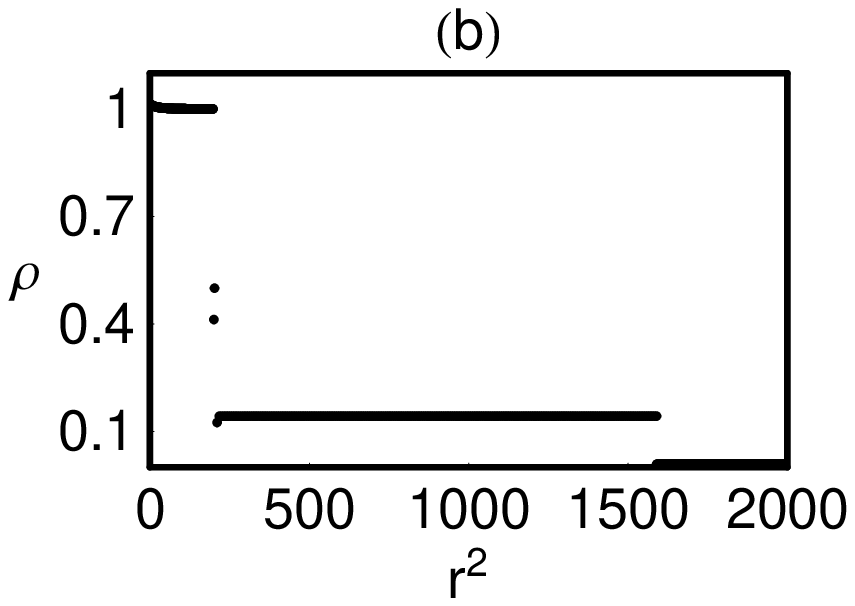}
\caption{Plot of the density for the generalized Jain states:
(a) $1/\nu_{cl}=1/k-1$ (\ref{generalizedm3});
(b) $1/\nu_{cl}=1/k-3/2$ (\ref{generalizedm4}), with $k=4$ and $N=400$.}
\end{figure}

The first non-trivial value is $n=2$, i.e. $m=3$ in Table 1:
\ba
B &=& \sum_{n=1}^{N/2} \sqrt{(k-2)n} \mid n\ra \la n-1 \mid
\ +\ \sum_{n=\frac{N}{2}+1}^{N-1} \sqrt{(k+2)n-2N} \mid n \ra \la n-1 \mid ,
\nl
A &=& \sum_{n=0}^{\frac{N}{2}-1} \sqrt{2} \mid \frac{N}{2}+n \ra \la n \mid .
\label{generalizedm3}
\ea
In matrix representation for $N=4$:
\ba
{\scriptsize
B = \left(\begin{array}{cccc} 
0 & 0 & 0 & 0 \\ \sqrt{k-2} & 0 &  0 & 0 \\ 
0 & \sqrt{2(k-2)} & 0 & 0 \\ 0 & 0 & \sqrt{3k-2} & 0 
\end{array}\right)\ \ , \qquad
A = \left(\begin{array}{cccc} 
 0 & 0 &  0 & 0 \\ 0 & 0 & 0 & 0 \\ 
\sqrt{2} & 0 & 0 & 0 \\ 0 & \sqrt{2} & 0 & 0 
\end{array}\right)\ } \ . \
\ea

The analogous state for $n=3$, corresponding to 
$(\Phi_{1/2, \ gs})^3 \ (\Phi_{1,\ gs})^{k-3}$, is:
\ba
B &=& \sum_{n=1}^{N/2} \sqrt{(k-3)n} \mid n \ra \la n-1 \mid\ +\ 
\sum_{n=\frac{N}{2}+1}^{N-1} \sqrt{(k+3)n-3N} \mid n \ra \la n-1 \mid ,
\nl
A &=& \sum_{n=0}^{\frac{N}{2}-1} \sqrt{3} \mid \frac{N}{2}+n \ra \la n \mid , \
\label{generalizedm4}
\ea
and in matrix form for $N=4$:
\ba
{\scriptsize
B = \left(\begin{array}{cccc} 
0 & 0 & 0 & 0 \\ \sqrt{k-3} & 0 &  0 & 0 \\ 
0 & \sqrt{2(k-3)} & 0 & 0 \\ 0 & 0 & \sqrt{3(k-1)} & 0 
\end{array}\right)\ \ , \qquad
A = \left(\begin{array}{cccc}  
0 & 0 &  0 & 0 \\ 0 & 0 & 0 & 0 \\ \sqrt{3} & 0 & 0 & 0 
\\ 0 & \sqrt{3} & 0 & 0 \end{array}\right)\ } \ . \
\ea
In Fig.5, we plot the density for these generalized Jain states: 
these are droplets with two-step constant density similar
to that of composite-fermion states. 
At present, we do not have strong arguments to dispose of 
these additional states: this issue will be further discussed 
in the conclusion.


\subsubsection{States with many-step density}

Other generalized matrix Jain states (\ref{ext-jain}) 
in Table 1, for $A^m\approx 0$, $m \ge 4$, are
made by the product of three or more 
different terms. The simplest one for $m=4$ is
$\Phi_\nu=\Phi_1^{k-2}\ \Phi_{1/2}\ \Phi_{1/3}$ with energy $E/\B=3N/2$ and
angular momentum $J\sim(k-1-1/6)N^2/2$.
Within the droplet interpretation of classical solutions discussed
before, we seek for a three-step solution ($N$ multiple of $6$),
$$
\b_i\sim i(k-2)\ ,\quad 1<i<\frac{N}{2}\ ;\qquad
\b_i\sim ik\ ,\quad \frac{N}{2}<i<\frac{2N}{3}\ ;\qquad
\b_i\sim i(k+3)\ ,\quad \frac{2N}{3}<i<N\ .
$$
However, there is no simple $A_{ab}$ solution  with
entries $(0,1,\sqrt{2},\sqrt{3})$, that fulfills the Gauss law equation
for the same energy and angular momentum of the quantum state.
We proved this fact for an ansatz with piecewise
constant density making up to $6$ steps.
A four-step solution (see Fig.6) can be found
with quantum numbers differing macroscopically from the quantum values,
$E\sim 1.4\ N, J\sim (k-1-0.14)N^2/2$: in particular, the larger angular
momentum identifies it as an excited state.
Presumably, the quantum state can be better approximated 
by allowing a very large number of steps, leading to a 
modulated (or singular) density profile in the large-$N$ limit.
This result indicates that most of the multi-component generalized 
matrix quantum states found in \cite{c-rod} for projections $A^m\ap 0$,
$m\ge 4$, are not semiclassical incompressible fluids.

\begin{figure}
\begin{center}
\includegraphics[width=8cm]{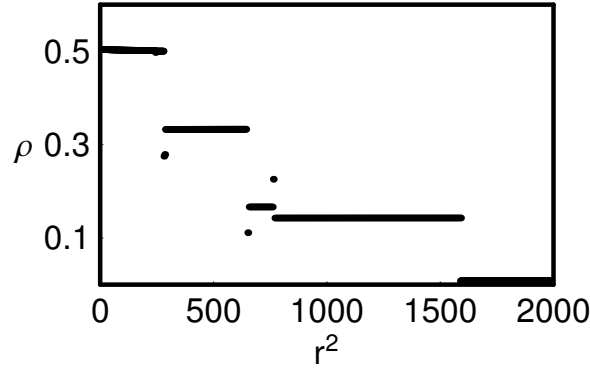}
\end{center}
\caption{Plot of the density of the 4-step excited state in the $A^4\ap 0$ 
theory, $1/\nu_{cl}\sim k-1-0.14$, with $k=4$ and $N=400$.}
\end{figure}


\subsection{Quasi-holes solutions}

As recalled in section 2.3, the $g=0$ matrix theory, projected
by $A^m\ap 0$, possess quasi-hole excitations above the $\nu^*=m$ 
Jain ground states.
In this section we give the corresponding semiclassical 
solutions for $\nu^*=2$, corresponding to deformation of the density
of solution (\ref{nu2-sol}) in fig.4(a). 

The classical equation of motion for $A$ and $B$, eq. (\ref{eq-a},\ref{eq-b}),
are linear and admit a general solution for excitations:
\ba
A_{ab}(t) &=& {\rm e}^{-i\left(\G_a+2\right)t}
\ \left({\rm e}^{it\L}\ A(0)\ {\rm e}^{-it\L}\right)_{ab}\  
{\rm e}^{-i\left(\G'_b\right)t}\ ,
\nl
B(t) &=& {\rm e}^{-i\w t}
\ {\rm e}^{it\L}\ B(0)\ {\rm e}^{-it\L}\ . 
\label{ext-eq}
\ea
Therefore, we should only solve the Gauss law (\ref{eq-g}) and
the constraint (\ref{eq-z}).

In the two-step fluid density in fig.4(a), 
one can have more than one quasi-hole corresponding
to punching either of the two possible fluids.
A hole in the complete fluid is obtained by generalizing
the quasi-hole of the Laughlin state found in ref.\cite{poly1}: it is a 
deformation of the $B$ matrix (\ref{nu2-sol}) that
describes a quasi-hole of charge $q>0$ situated at the origin. The
matrix $A$ remains unchanged:
\ba
B &=& \sum_{n=1}^{N/2} \sqrt{(k-1)(n+q)} \mid n \ra \la n-1 \mid 
\nl
&&+\ \sum_{n=\frac{N}{2}+1}^{N-1} 
\sqrt{(k-1)q + n(k+1) - N} \mid n \ra \la n-1 \mid \ + \ 
\sqrt{(k-1)q} \mid 0 \ra \la N-1 \mid \ ,
\nl
A &=& \sum_{n=0}^{\frac{N}{2}-1} \mid n+ \frac{N}{2} \ra \la n \mid. \
\label{qhorigin_nu2}
\ea
In matrix representation:
\ba
\!\!\! B = {\scriptsize\left(\begin{array}{cccc} 
0 & 0 & 0 & \sqrt{q(k-1)} \\ \sqrt{(1+q)(k-1)} & 0 &  0 & 0 
\\ 0 & \sqrt{(2+q)(k-1)} & 0 & 0 \\ 0 & 0 & \sqrt{(3+q)k-1-q} & 0 
\end{array}\right),}\quad
A = {\scriptsize\left(\begin{array}{cccc}  
0 & 0 &  0 & 0 \\ 0 & 0 & 0 & 0 \\ 1 & 0 & 0 & 0 \\ 0 & 1 & 0 & 0 
\end{array}\right).}
\ea
The corresponding density is shown in Fig.7(a).

\begin{figure}
\begin{center}
\includegraphics[width=9cm]{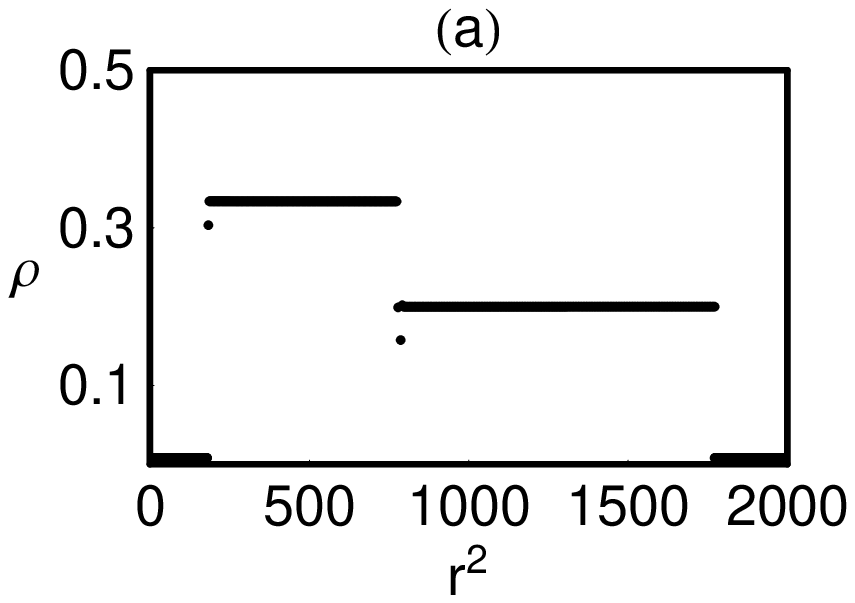}

\includegraphics[width=9cm]{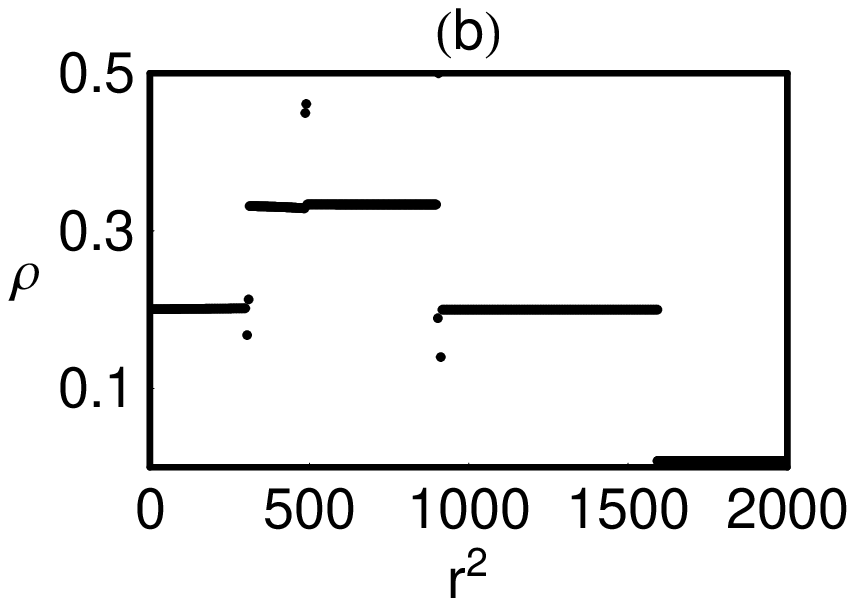}

\includegraphics[width=9cm]{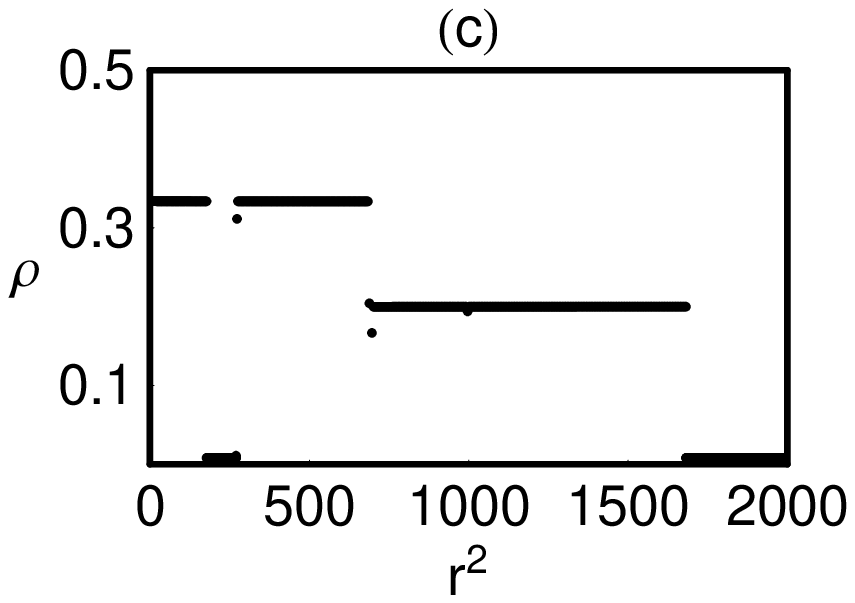}
\end{center}
\caption{Plot of the density of the $\nu^*=2$ Jain ground state,
$1/\nu_{cl}=k+1/2$,  for $k=4$ and $N=400$, including: 
(a) one quasi-hole in the origin (\ref{qhorigin_nu2}) with $q=60$; 
(b) a quasi-hole in the upper layer of the fluid (\ref{qh2steps_nu2}) 
with $q=60$; 
(c) the quasi-hole out of the origin (\ref{qh2steps3_nu2}) with $q=30$
and $r=60$.}
\end{figure}

A quasi-hole only affecting the upper layer of the $\nu^*=2$ fluid is shown in 
Fig.7(b). It is given by the solution:
\ba
B &=& 
\sum_{n=0}^{q} \sqrt{(k+1)(n+1)} \mid n+1 \rangle \langle n \mid + 
\sum_{n=q+1}^{\frac{N}{2}+q} 
\sqrt{(k-1)(\a-q-1+n)} \mid{n+1} \rangle \langle n \mid 
\nl
&&+\ \sum_{n=\frac{N}{2} + q + 1} ^ {N-2} 
\sqrt{(k+1)(\b + n - \frac{N}{2} -q -1)} \mid n+1 \rangle \langle n \mid ,
\nl
A &=& \sum_{n=0}^{q} \mid n \rangle \langle q + 1 + n \mid + 
\sum_{n=0} ^ {\frac{N}{2} -q -2} 
\mid \frac{N}{2} + q + 1 + n \rangle \langle 2q + 2 +n \mid \ , \
\label{qh2steps_nu2}
\ea
with $\a= \frac{q + k(2+q)}{k-1}$, 
$\b=\frac{2+q+(k-1) \frac{N}{2}}{k+1}$ and $q$ a positive integer.
In matrix representation for $N=8$ and $q=1$, it reads:
\ba
B &=& {\scriptsize \left(\begin{array}{cccccccc} 
0 & 0 & 0 & 0 & 0 & 0 & 0 & 0\\ 
\sqrt{(k+1)} & 0 &  0 & 0 & 0 & 0& 0 & 0\\ 
0 & \sqrt{2(k+1)} & 0 & 0 & 0 & 0& 0 & 0 \\ 
0 & 0 & \sqrt{3k+1} & 0 & 0 & 0& 0 & 0 \\ 
0 & 0 & 0 & \sqrt{4k} & 0 & 0 & 0 & 0\\ 
0 & 0 & 0 & 0 & \sqrt{5k-1} & 0 & 0 & 0\\
0 & 0 & 0 & 0 & 0 & \sqrt{6k-2} & 0 & 0 \\ 
0 & 0 & 0 & 0 & 0 & 0 & \sqrt{7k-1} & 0 
\end{array}\right)} \ , \nl
A &=& {\scriptsize \left(\begin{array}{cccccccc}  
0 & 0 & 1 & 0 & 0 & 0 & 0 & 0\\ 
0 & 0 & 0 & 1 & 0 & 0 & 0 & 0\\ 
0 & 0 & 0 & 0 & 0 & 0 & 0 & 0\\ 
0 & 0 & 0 & 0 & 0 & 0 & 0 & 0\\
0 & 0 & 0 & 0 & 0 & 0 & 0 & 0\\
0 & 0 & 0 & 0 & 0 & 0 & 0 & 0\\ 
0 & 0 & 0 & 0 & 1 & 0 & 0 & 0\\
0 & 0 & 0 & 0 & 0 & 1 & 0 & 0 
\end{array}\right).}\
\ea
The displacement from the origin of the upper layer 
corresponds to $\D {\cal J}=(k+1)(q+1)$.

It is also possible to create a circular depletion in the whole fluid,
of size (charge) $\D{\cal J}=q$ outside the origin at a distance
$\D {\cal J}=(k-1)(r-1)$ (see fig.7(c)): 
\ba
B &=& \sum_{n=0}^{r-1} \sqrt{(k-1)(n+1)} \mid n+1 \rangle \langle n \mid + 
\sum_{n=r}^{\frac{N}{2}-2} \sqrt{(k-1)(n+1+q)} 
\mid{n+1} \rangle \langle n \mid
\nl
&&+\ \sqrt{(k-1)q} \mid r \rangle \langle n-1 \mid + 
\sqrt{(\frac{N}{2} + q)(k-1)} \mid \frac{N}{2} \ra \la \frac{N}{2} - 1 \mid
\nl
&&+\ \sum_{n=\frac{N}{2}} ^ {N-2} 
\sqrt{(k+1)(n +1) - N + (k-1)q} \mid n+1 \rangle \langle n \mid ,
\nl
A &=& \sum_{n=0}^{\frac{N}{2}-1} 
\mid n + \frac{N}{2} \rangle \langle n \mid .
\label{qh2steps3_nu2}
\ea
In matrix representation for $N=6$ and $q=2$ :
\ba
B &=& {\scriptsize \left(\begin{array}{cccccc} 
0 & 0 & 0 & 0 & 0 & 0 \\ \sqrt{(k-1)} & 0 &  0 & 0 & 0 & 0\\
 0 & \sqrt{2(k-1)} & 0 & 0 & 0 & \sqrt{r(k-1)} \\ 
0 & 0 & \sqrt{(3+r)(k-1)} & 0 & 0 & 0 \\ 
0 & 0 & 0 & \sqrt{(4+r)k-2-r} & 0 & 0 \\ 
0 & 0 & 0 & 0 & \sqrt{(5+r)k-1-r} & 0 
\end{array}\right)\ ,} \qquad
\nl
A &=& {\scriptsize \left(\begin{array}{cccccc}  0 & 0 &  0 & 0 & 0 & 0 \\ 
0 & 0 & 0 & 0 & 0 & 0 \\ 0 & 0 & 0 & 0 & 0 & 0 \\ 
1 & 0 & 0 & 0 & 0 & 0 \\0 & 1 & 0 & 0 & 0 & 0 \\
0 & 0 & 1 & 0 & 0 & 0  
\end{array}\right).} 
\label{example}
\ea
In this case, the solution of the Gauss law is obtained in terms of a
two-component auxiliary field $\psi$, 
and it holds for $rq \ll N$, namely the depletion would move back to the
origin in the scaling limit $N\to\infty$.


\section{Conclusion}

In this paper, we have continued the study of 
the Maxwell-Chern-Simons matrix gauge theory as an
effective theory of quantum Hall states. After providing better
forms for the projection, $A^m\ap 0$, limiting state degeneracy, we
have obtained the semiclassical ground states of the theory.
They correspond to the quantum states found before \cite{c-rod}, that
reproduce the Jain composite-fermion construction of phenomenological
wave functions.
The density of states in the main Jain series, $\nu=m/(mk+1)$, has been found
to be that of incompressible fluids: this confirms our
expectation that the matrix states at $g=0$ are not too different
from the physical states at $g=\infty$.
The semiclassical approximation used here is known to be valid
in the large $N$ limit for both matrix \cite{mehta}\cite{wiegmann} 
and real Hall states \cite{ctz-class}\cite{sakita}, in 
particular for incompressible fluid states.

Outside the main series of hierarchical states,
other ground states are possible in the matrix theory \cite{c-rod}, 
that correspond to generalized Jain constructions \cite{jain}.
In Jain's theory, these generalized  states 
are excluded due to their low (or vanishing) gap.
In the semiclassical analysis of the matrix theory, we have found that
the majority of generalized states do not have piecewise constant density, 
i.e. are not incompressible fluids: this is an indication that they
may become unstable for finite $g>0$ values.

The study of the phase diagram of the matrix theory is clearly necessary
to make better contact between the nice results ($g=0$)  and the
physical regime ($g=\infty$), upon varying the potential
$V=g\ \Tr[\ov{X},X]^2\ $. We expect that the relevant incompressible
fluid states have a smooth evolution for $g>0$ and we plan to 
include the quartic potential in the semiclassical analysis
by means of a mean-field approximation.

The explicit semiclassical solutions in this paper
can also be useful to study the symmetries and algebraic properties
of matrix ground states. We would like:
\begin{itemize}
\item
To make contact with the $SU(m)$ symmetry of the conformal field theories
describing the edge excitations of Jain states \cite{winf}.
\item
To find a projection of states more refined than $A^m\ap 0$, that could
discriminate the hierarchical Jain states from the generalized (unstable) ones.
Such an expectation is based on the general belief that the 
observed Hall states should be uniquely characterized by algebraic 
conditions and gauge invariance, rather than by detailed dynamics, because
they are exceptionally robust and universal.
\end{itemize}

\bigskip

{\large \bf Acknowledgments}

We thank F. Colomo and G. R. Zemba for interesting discussions.
I. D. Rodriguez thanks the EC program Alban of Ph-D scholarships for
Latin American students.
This work was partially funded by the ESF programme 
``INSTANS: Interdisciplinary Statistical and Field Theory Approaches 
to Nanophysics and Low Dimensional Systems''.

\appendix


\section{Appendix}

\subsection{Gauge invariance of the projection}

Here is an explicit proof that the projection 
$(A_{ab})^m \Psi$ (cf \ref{sll-cond}) is a
gauge invariant condition on quantum states.
Consider the more general relation for $m=2$:
\be
A_{ab}\ A_{a',b'}\ \Psi\left( \ov{A} ,\ov{B} \right) \ = \ 
M_{bb'}\left( \ov{A} ,\ov{B} \right)\ V_a W_{a'}\ .
\label{cond-gen}
\ee
The wave function is assumed to be gauge
invariant: $\Psi(\ov{A},\ov{B})=\Psi(U\ov{A} U^\dag,U\ov{B} U^\dag)$.
The form in the r.h.s. is specific of the 
bush states of section 2, but this is not relevant for the argument.
The matrix $M_{bb'}$ vanishes for $a=a',b=b'$ because $\Psi$ is assumed to be
one solution of the constraint. In general, there are several
terms in the r.h.s. with that structure, but the matrices $M_{bb'}$ 
should all vanish independently because they are multiplied by 
monomials $V_a W_{a'}$ that are all independent
\cite{c-rod}.

Let us now multiply by unitary matrices and sum over indices on both
sides to realize a gauge transformation of the two $A$'s:
\ba
\left(U A U^\dag\right)_{ab}\ 
\left(UA U^\dag\right)_{a',b'}\ \Psi\left( \ov{A} ,\ov{B} \right) 
& = &
M_{\wt{b}\wt{b'}}\left( \ov{A} ,\ov{B} \right)\ 
U^\dag_{\wt{b}b}\ U^\dag_{\wt{b'}b'}\ (UV)_a (UW)_{a'}\ ,\nl
& = &
M_{bb'}\left(U \ov{A} U^\dag ,U\ov{B} U^\dag \right)\ 
 (UV)_a (UW)_{a'}\ .
\label{cond-proof}
\ea
The resulting expression of $M(U\ov{A} U^\dag,U\ov{B} U^\dag)_{bb'}$
vanishes whenever $M(\ov{A},\ov{B})_{bb'}$ does, i.e. for $b=b'$,
because both vanish by polynomial identities that do not
depend on the specific vales of the variables.
Therefore, a solution of the constraint remains a solution after
gauge transformation.


\end{document}